\documentclass[onecolumn,authoryear]{els-mrw} 

\usepackage{amsmath,amssymb,amsfonts,amsthm,makeidx,graphicx}
\usepackage{txfonts}
\usepackage{helvet}
\usepackage{comment}
\usepackage[colorlinks]{hyperref}
\hypersetup{linkcolor=black, citecolor=black, urlcolor=black}
\usepackage{mathtools} 
\usepackage{scalerel}[2016-12-29]
\usepackage{multirow}

\def\stretchint#1{\vcenter{\hbox{\stretchto[440]{\displaystyle\int}{#1}}}}

\begin{document}

\chapter{Dark energy and cosmic acceleration}\label{chap1}

\author[1,2]{Rodrigo von Marttens}%
\author[3]{Jailson Alcaniz}%


\address[1]{\orgname{Universidade Federal da Bahia}, \orgdiv{Instituto de Física}, \orgaddress{Salvador - BA, 40170-110, Brasil}}
\address[2]{\orgname{Universidade Federal do Espírito Santo}, \orgdiv{PPGCosmo}, \orgaddress{Vitória - ES, 29075-910, Brasil}}
\address[3]{\orgname{Observatório Nacional}, \orgaddress{Rio de Janeiro - RJ, 20921-400, Brasil}}

\articletag{Chapter Article tagline: update of previous edition,, reprint..}

\maketitle

\begin{glossary}[Nomenclature]
    \begin{tabular}{@{}lp{34pc}@{}}
    GR & General Relativity \\
    DE & Dark Energy \\
    CDM & Cold Dark Matter \\
    CP & Cosmological Principle \\
    SN Ia & Type Ia Supernovae \\
    CMB & Cosmic Microwave Background \\
    BAO & Baryon Acoustic Oscilation \\
    DETF & Dark Energy Task Force \\
    \end{tabular}
    \end{glossary}
    
    \begin{abstract}
    The discovery that we live in an accelerating universe changed drastically the paradigm of physics and introduced the concept of \textit{dark energy}. In this work, we present a brief historical description of the main events related to the discovery of cosmic acceleration and the basic elements of theoretical and observational aspects of dark energy. Regarding the historical perspective, we outline some of the key milestones for tracing the journey from Einstein's proposal of the cosmological constant to the type Ia supernovae results. Conversely, on the theoretical/observational side, we begin by analyzing cosmic acceleration within the context of the standard cosmological model, i.e., in terms of the cosmological constant. In this case, we show how a positive cosmological constant drives accelerated expansion and discuss the main observational aspects, such as updated results and current cosmological tensions. We also explore alternative descriptions of dark energy, encompassing dynamic and interacting dark energy models.
    \end{abstract}

    \section{Introduction}\label{sec:intro}

    Dark energy (from now on, DE) has emerged as a theoretical concept built upon empirical observations, indicating that our Universe is undergoing an accelerated phase of expansion~\citep{Perlmutter:1998np,Riess:1998cb}. This cosmic acceleration may appear paradoxical given the conventional understanding of the attractive nature of gravity. This means that something beyond ``usual physics'' is required to explain the underlying mechanism behind the acceleration of the universe~\citep{Uzan:2006mf}. Understanding the genuine nature of DE is a major challenge in modern cosmology. In the coming years, a substantial allocation of resources and human power is expected for this purpose~\citep{Slosar:2019flp,DiValentino:2020vhf}.
    
    From a formal perspective, the accelerated expansion of the Universe can be reached by incorporating an effective negative pressure into its dynamical framework~\citep{amendola2010dark}. The standard approach introduces a positive cosmological constant into Einstein's equations, typically symbolized by $\Lambda$. At first sight, describing DE via the cosmological constant is a convenient choice, because it can be easily incorporated into the general relativity formulation and naturally aligns with the vacuum energy. Nevertheless, a considerable disparity exists between the theoretical predictions for vacuum energy from quantum field theory and observational data, differing by over a hundred orders of magnitude. This huge inconsistency is commonly called the cosmological constant problem~\citep{Weinberg:1988cp,Martin:2012bt,Lombriser:2019jia}. Despite the unresolved connection between the cosmological constant and vacuum energy, describing the DE through the cosmological constant remains the standard approach. 
    
    Within the framework of General Relativity, the combination of the cosmological constant and Cold Dark Matter (CDM) constitutes the so-called dark sector of the universe, which constitutes about 95\% of the current energy content of the cosmic substratum. This dark sector serves as one of the foundations of the standard cosmological model, also known as the $\Lambda$CDM model ($\Lambda$ representing the cosmological constant and CDM representing cold dark matter). The standard cosmological model delivers a comprehensive description of the late-time evolution of the universe, including the ongoing phase of accelerated expansion. An impressive achievement of the $\Lambda$CDM model lies in its ability to accurately model a wide range of cosmological observables, successfully fitting their corresponding observational data while employing the same set of values for the cosmological parameters~\citep{Planck:2018vyg,eBOSS:2020yzd,Brout:2022vxf,DES:2021wwk}. 
    
    Despite its simplicity and considerable observational success, the standard cosmological model still faces important open questions from both theoretical and observational perspectives~\citep{Perivolaropoulos:2021jda}, many of which are related to topics around the nature of the DE and its connection with the cosmological constant (cf.~\citet{Bianchi:2010uw}). Regarding the theoretical side, besides the aforementioned cosmological constant problem, it is intriguing the fact that, despite their independent time evolutions, the energy density associated with the cosmological constant has only become relevant in recent epochs. This question ``why now?'' is often referred to in the literature as the Cosmic Coincidence Problem~\citep{steinhardtcritical,Weinberg:2000yb,Velten:2014nra}. On the observational side, recent observations have shown a significant deviation from the $\Lambda$CDM predictions derived from the Planck 2018 analysis~\citep{Planck:2018vyg}\footnote{Henceforth, references to the Planck 2018 analysis denote the TTTEEE+lensing+BAO results, presented in the last column of Table 2 in~\citet{Planck:2018vyg}.}. The most notable discrepancies include the determination of the amplitude of matter perturbations, and the measurement of the current value of the Hubble parameter. Whereas in the first case, weak lensing results appear to yield a deviation of about $3\sigma$~\citep{KiDS:2020suj}, the second case is more problematic, with a discrepancy of more than $4\sigma$~\citep{Riess:2021jrx}.
    
    All these aspects have been, and continue to be, invoked as motivation for exploring alternative descriptions of DE beyond the cosmological constant. The most widely adopted approaches for seeking alternative descriptions for the accelerated expansion of the universe are: ($i$) Introducing an additional exotic energy component to the cosmic substratum. Given our lack of knowledge regarding the nature of DE, this extra component is commonly introduced as a generic scalar field under the influence of a given potential. At cosmological scales, this scalar field can typically be described as a fluid with a negative equation of state. ($ii$) Consider the proposition that General Relativity may not be the most appropriate theory for describing gravity on large scales. In such a scenario, General Relativity acts as an effective description, valid for typical solar system scales. However, to reproduce the observed accelerated expansion, gravity would undergo significant modifications on larger scales. A particular scenario where a modified gravity theory is written in terms of both a metric and a scalar field is known as scalar-tensor theories~\citep{faraoni2004scalar}\footnote{A historically significant example is the Brans-Dicke theory~\citep{brans1961mach,dicke1962mach}.}. 
    
    In general, the term ``dark energy model'' is typically associated with the first approach, while the second approach is referred to as modified gravity. Restricting ourselves to DE models, the simplest case consists in considering a scalar field with canonical kinetic term. This approach is commonly addressed as quintessence models~\citep{Fujii:1982ms,Ratra:1987rm,Tsujikawa:2013fta}. Each specific model is defined by the chosen potential~\citep{Copeland:1997et,Caldwell:1997ii}. In order to satisfy the acceleration condition, the scalar field must meet conditions equivalent to the slow-roll criteria applied to the inflation period. This approach can be generalized to the case with a general kinetic term, often referred to as $k$-essence models~\citep{Chiba:1999ka,Armendariz-Picon:2000nqq}. In this case, the DE comoving sound speed may deviate from the speed of light, potentially leading to observable clustering effects of DE~\citep{Batista:2021uhb}. The scalar-field approach can be consistently written as an effective fluid whose equation of state is determined by the potential. Although equivalent, the fluid approach may appear to be theoretically simpler and more straightforward for interpreting the dynamics. Considering this equivalence, along with the simplicity of the fluid approach, it is common for DE models to be motivated directly as a fluid. In such cases, instead of specifying a potential, an equation of state for DE is chosen. For example, as a direct extension of the cosmological constant case, a usual approach consists of to consider DE as a fluid with a constant equation of state, which need not necessarily be -1. This methodology is commonly known as $w$CDM. Additionally, more complex scenarios can also be considered, where the equation of state of DE is time-dependent. Another case of great interest is the Chevallier-Polarski-Linder (CPL) parameterization~\citep{Chevallier:2000qy,Linder:2002et}. The CPL parameterization has been considered in the Dark Energy Task Force (DETF) figure of merit~\citep{Albrecht:2006um}.
    
    Another possibility involves considering that DE exhibits a non-minimal coupling with matter~\citep{Billyard:2000bh,Amendola:1999er,Zimdahl:2001ar}. This scenario can be interpreted as a non-gravitational interaction between DE and another material component. A common choice is to explore the interactions within the dark sector~\citep{Majerotto:2009zz,vonMarttens:2018iav}. In this framework, a source term is introduced into the conservation equations of the interacting components, which results in an energy/momentum exchange. This choice is motivated by our limited understanding of the dark components and also because they constitute the dominant energy content of the present Universe. Initially introduced to address or alleviate the cosmic coincidence problem~\citep{Chimento:2003iea}, these models are now also being utilized as an alternative approach to address current tensions in cosmology~\citep{DiValentino:2019jae}.
    
    Recent developments explore scenarios that combine features from DE models with modified gravity theories. In this context, an interesting approach consists in effective field theory for DE~\citep{Gubitosi:2012hu,Bellini:2014fua,Gleyzes:2014rba}. This methodology considers the most general gravity action, with the aim of understanding the effects preferred by observational data~\citep{Gleyzes:2013ooa}. Specifically, focusing on the Horndeski picture~\citep{Horndeski:1974wa}, i.e., the most general local, Lorentz-covariant, and four-dimensional scalar-tensor gravity where the Euler-Lagrange equations are, at most, second-order in the derivatives of the scalar and tensor field, it becomes feasible to distinguish between terms associated with DE clustering on both large and small scales (effects connected to DE models) and terms related to variations in Planck mass or coupling of kinetic terms with gravity (effects connected to modified gravity theories).
    
    In this context, the main goal of this text is to present and discuss the basic concepts related to the accelerated expansion of the universe, focusing on the scenario of the DE models. Given the importance of this topic, there are numerous high-quality review articles~\citep{Peebles:2002gy,Padmanabhan:2002ji,Sahni:2004ai,Copeland:2006wr,Durrer:2011gq,Mortonson:2013zfa,Huterer:2017buf} and cosmology textbooks~\citep{weinberg2008cosmology,liddle2015introduction,dodelson2020modern,baumann2022cosmology,huterer2023course} that have addressed the cosmic acceleration from different perspectives. Among them, it is worth highlighting the textbook~\citep{amendola2010dark}, which is focused on DE. This work is organized in the following way: in Sect.~\ref{sec:hist}, a brief historical description of the main developments leading to the discovery of the accelerated expansion of the universe; in Sect.~\ref{sec:lcdm} the mathematical basis of the $\Lambda$CDM model is presented, and some important observational aspects are discussed; Sect.~\ref{sec:models} is intended to address some of the most popular alternative DE models; and the final conclusions are presented in Sect.~\ref{sec:conclusions}.
    
    \section{A brief historical perspective}\label{sec:hist}

    The history of the developments related to the description of the accelerated expansion of the Universe is deeply connected to advancements in calculating distances and recession velocities, as well as to the history of the cosmological constant. It is important to mention that the first measurements of the radial velocities of galaxies were carried out in 1912,~\citep{slipher1913radial}, even before Einstein published the theory of general relativity. The cosmological constant first appeared in~\citet{Einstein:1917ce}. In this work, Einstein introduced it as a mathematical tool to yield a solution for a static universe. It is indeed ironic that the term introduced to maintain the universe static emerged as the main candidate to explain its accelerated expansion. Furthermore, it is worth mentioning that the proposed solution was instable and there was no association of the cosmological constant with the vacuum energy. This identification occurred five decades later in~\citet{Zeldovich:1967gd}.
    
    Years later Einstein's paper, Alexander Friedmann proposed a novel cosmological solution in which the universe expands~\citep{Friedman:1922kd}. However, due to a positive curvature (closed universe), it ultimately achieves a maximum radius before collapsing in the future. This work marked the first description of an expanding Universe within the context of General Relativity. The case with negative curvature was also explored by Friedmann in subsequent years~\citep{Friedmann:1924bb}. Shortly thereafter, Einstein published a note referring to~\citet{Friedman:1922kd} as an error. However, he was later convinced to recognize the validity of Friedmann's solution, which motivated him to publish a retraction. The complete translation of the notes published by Einstein are reproduced in the following box\footnote{Both transcriptions can be found in~\cite{1987Ap&SS.134..422B}.}.
    \begin{BoxTypeA}[chap1:box1]{Transcription of the Einstein's notes in reference to~\citet{Friedman:1922kd}}
    \section*{First Note:}
    \begin{center}
        Remark on the work of A. Friedmann~\citep{Friedman:1922kd}\\
        ``On the curvature of space''\\
        A. Einstein, Berlin\\
        Received September 18, 1922\\
        Zeitschrift f\u{u}r Physik
    \end{center}
    
    The work cited contain a result concerning a non-stationary world which seems suspected to me. Indeed, those solutions do not appear compatible with the field equations (A). From the field equations it follows necessarily that the divergence of the matter tensor $T_{ik}$ vanishes. This along with the anzatzes (C) and (D) leads to the condition
    \begin{eqnarray}
        \frac{\partial\rho}{\partial x_4}=0 \nonumber
    \end{eqnarray}
    which together with (8) implies that the world-radius $R$ is constant in time. The significance of the work therefore is to demonstrate this constancy.
    \vspace{0.5cm}
    \subsection*{Second note:}
    \begin{center}
        A note on the work of A. Friedmann\\
        ``On the curvature of space''\\
        A. Einstein, Berlin\\
        Received May 31, 1923\\
        Zeitschrift f\u{u}r Physik
    \end{center}
    
    I have in an early note (Einstein 1922) criticized the cited work~\citep{Friedman:1922kd}. My objection rested however -- as Mr. Krutko in person and a letter from Mr. Friedmann convinced me -- on a calculational error. I am convinced that Mr. Friedmann's results are both correct and clarifying. They show that in addition to the static solutions to the field equations there are time varying solutions with a spatially symmetric structure.
    \end{BoxTypeA}
    
\begin{figure}[t]
    \centering
    \includegraphics[width=.27\textwidth]{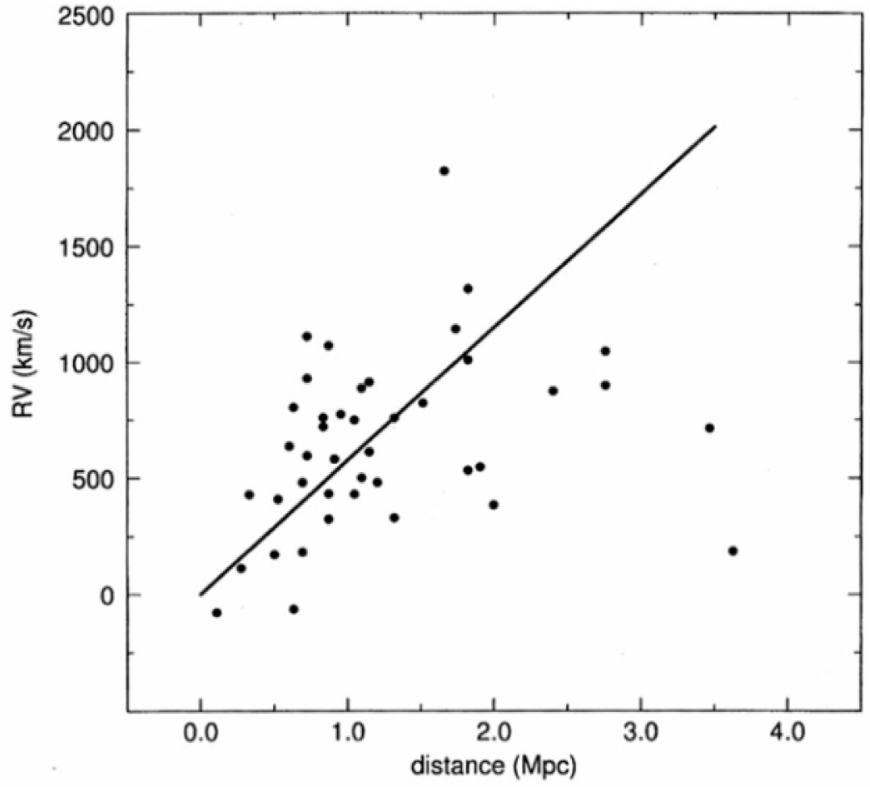}
    \includegraphics[width=.4\textwidth]{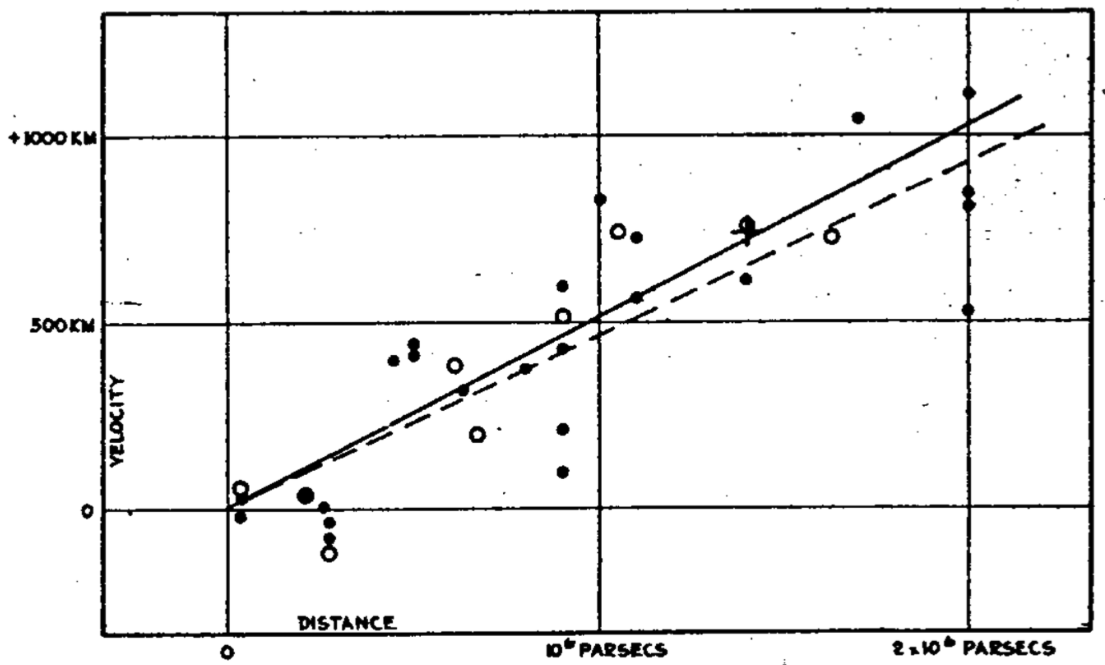}
    \includegraphics[width=.31\textwidth]{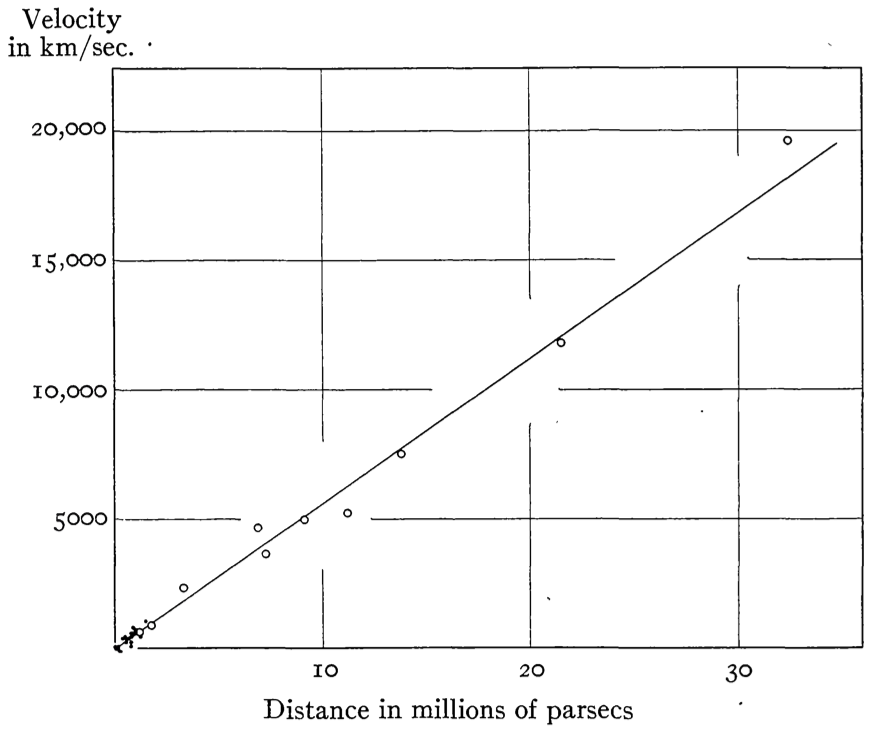}
    \caption{Velocity-distance diagrams used by Georges Lemaître and Edwin Hubble. These results have been the basis for the conclusion that our universe is expanding. As time passed, a better understanding of the systematic uncertainties inherent in measuring galaxy movements increased the robustness of the data. \textbf{Left panel: }data used in~\citet{Lemaitre:1927zz}. \textbf{Central panel: }data used in~\citet{Hubble:1929ig}. \textbf{Right panel: }data used in~\citet{Hubble:1931zz}. \\
    \textit{Source:} \citet{Lemaitre:1927zz,Hubble:1929ig,Hubble:1931zz}.}
    \label{fig:distvsvel}
    \end{figure}

    The observational validation through measurements of galaxy recession velocities was subsequently achieved through the contributions of Georges Lemaître~\citep{Lemaitre:1927zz} and Edwin Hubble~\citep{Hubble:1929ig,Hubble:1931zz}. Although the data used in Lemaître's analysis did not correct for peculiar motions, Hubble's data accounted for this effect. Consequently, as depicted in Fig.\ref{fig:distvsvel}, the credit for the observational verification of the expansion of the universe is more closely associated with Hubble. Nevertheless, Lemaître made significant contributions to the development of what has been known for decades as Hubble's law. Currently, in acknowledgment of his significant contributions, Hubble's law has been renamed the Hubble-Lemaître law\footnote{The final resolution of the International Astronomical Union (IAU) suggesting the renaming of the Hubble law to Hubble-Lemaître law can be found in~\href{https://www.iau.org/static/archives/announcements/pdf/ann18029e.pdf}{https://www.iau.org/static/archives/announcements/pdf/ann18029e.pdf}.}. 

    This result motivated the famous Einstein declaration, stating that the cosmological constant had been the biggest blunder of his life. Another funny aspect of this discovery is that the value obtained by~\citet{Hubble:1931zz} for the slope of the velocity-distance diagram, known as the Hubble constant, was $H_0=558\text{ km s}^{-1} {\rm Mpc}^{-1}$, which is very far from the current values. Additionally, it is crucial to emphasize that while this finding confirmed the expansion of the Universe, the realization of its acceleration only occurred years later.
    
\begin{figure}[t]
    \centering
    \includegraphics[width=.33\textwidth]{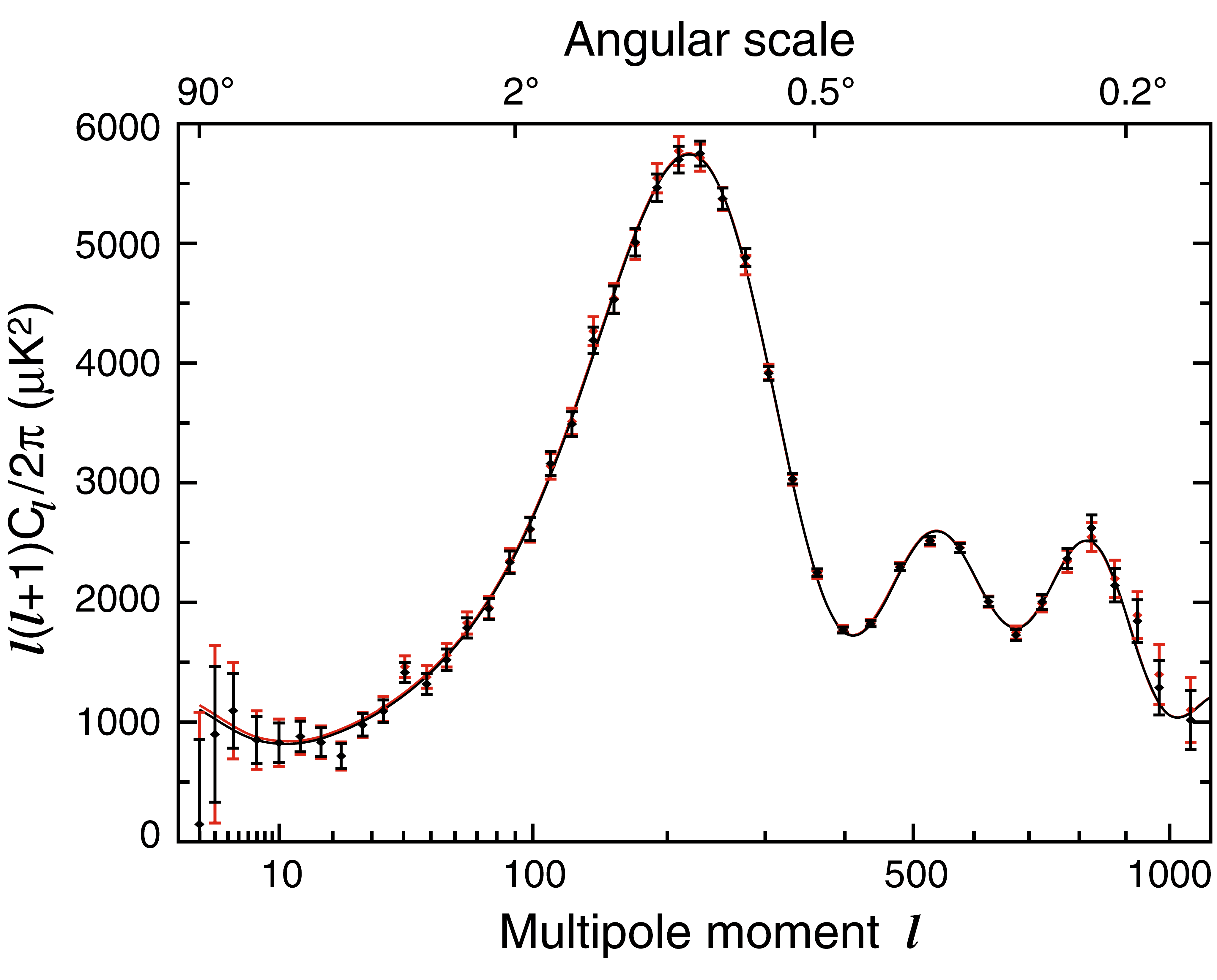} 
    \includegraphics[width=.65\textwidth]{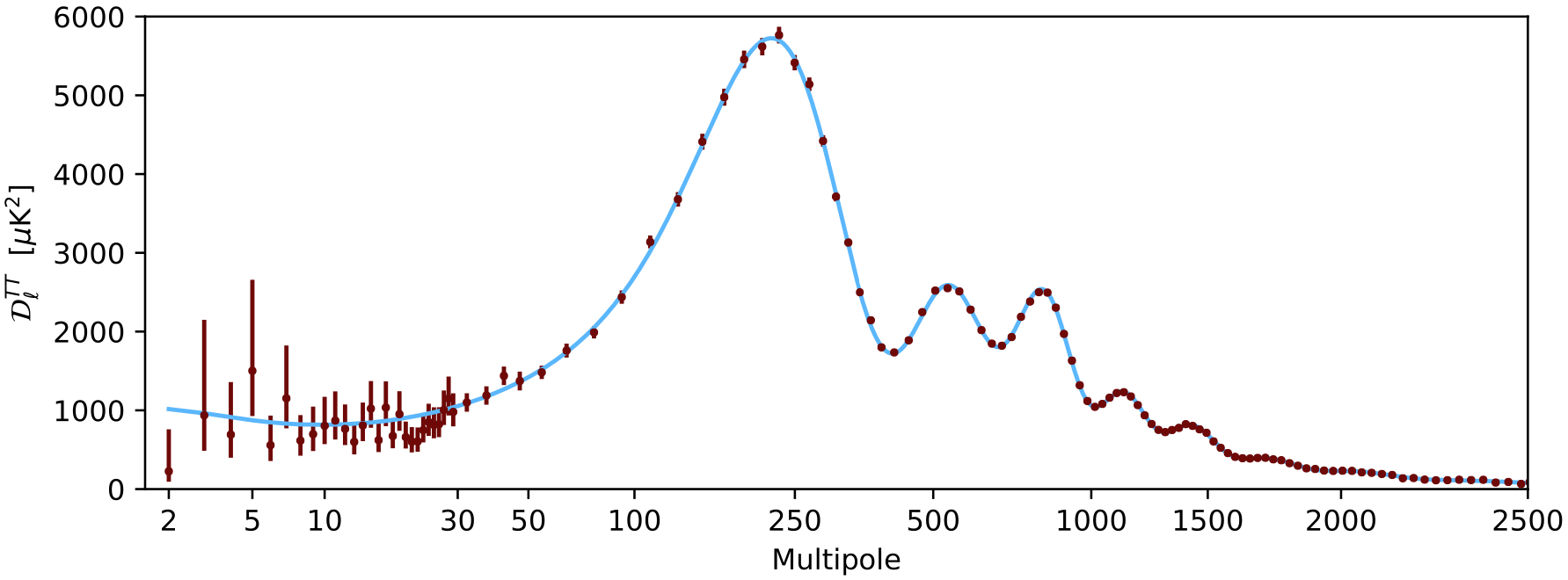}
    \caption{Angular power spectrum of the CMB temperature anisotropies. \textbf{Left panel: }observational result obtained by WMAP. \textbf{Right panel: }observational result obtained by Planck satelite. \\
    \textit{Source:} \citet{WMAP:2012nax,Planck:2018vyg}.}
    \label{fig:cmb}
    \end{figure}

    In the following decades, cosmology experienced significant advancements. Among others, one can highlight the following milestones: the proposition of the existence of CDM~\citep{Zwicky:1933gu}, and its subsequent finding in galaxies~\citep{Rubin:1970zza}; the prediction of primordial nucleosynthesis~\citep{Gamow:1946eb,Alpher:1948ve}; the discovery of the Cosmic Microwave Background (CMB)~\citep{Penzias:1965wn}; and the proposal of a primordial inflationary epoch~\citep{Guth:1980zm}. All of these developments delineate the evolution of modern cosmology. One of the most important aspects that significantly contributed to the observational validation of these proposals is the analysis of the CMB data. In the early 1990s, the first mission dedicated to mapping the CMB was launched: the Cosmic Background Explorer (COBE). This mission demonstrated excellent compatibility of the CMB temperature spectrum with a blackbody spectrum and also determined its temperature~\citep{Fixsen:1996nj}. Subsequent missions such as the Wilkinson Microwave Anisotropy Probe (WMAP) and the Planck satellite furthered this analysis by measuring the anisotropies present in the CMB~\citep{WMAP:2012nax,Planck:2018vyg}. The CMB data continue to have significant relevance for precision cosmology. To illustrate the relevance of CMB data in current observational cosmology, the final Planck 2018 analysis, in the context of $\Lambda$CDM model, achieved an unprecedented level of precision in determining cosmological parameters, surpassing 1\% threshold. The remarkable agreement between the $\Lambda$CDM model and the observational data from WMAP and Planck is depicted in Fig.~\ref{fig:cmb}.

    In the meantime, between the COBE and WMAP missions, another CMB experiment called Balloon Observations of Millimetric Extragalactic Radiation and Geomagnetics (BOOMERanG) successfully captured the first peak of the CMB spectrum~\citep{Boomerang:2000efg}. This finding led to the conclusion that the Universe exhibits spatial flatness. On the other hand, the analysis of large-scale structures in the Universe revealed that pressureless matter should constitute only 30\% of the total energy content of the Universe~\citep{Trimble:1987ee,Carlberg:1995aq,Bahcall:1997ia}. Consequently, reconciling these results would require identifying the source of the remaining 70\% of the cosmic energy content. In this context, even before the discovery of the accelerated expansion of the Universe, several works began to speculate about the potential possibility that the cosmological constant could account for the remaining energy content of the Universe~\citep{Efstathiou:1990xe,Ostriker:1995su}. Observational data from the ages of astrophysical objects, such as globular clusters and white dwarfs, were also employed to support the reconsideration of the cosmological constant in the dynamics of the Universe~\citep{Krauss:1995yb,Alcaniz:1999kr,Lima:2001zz}.
    
\begin{figure}[t]
    \centering
    \includegraphics[width=.51\textwidth]{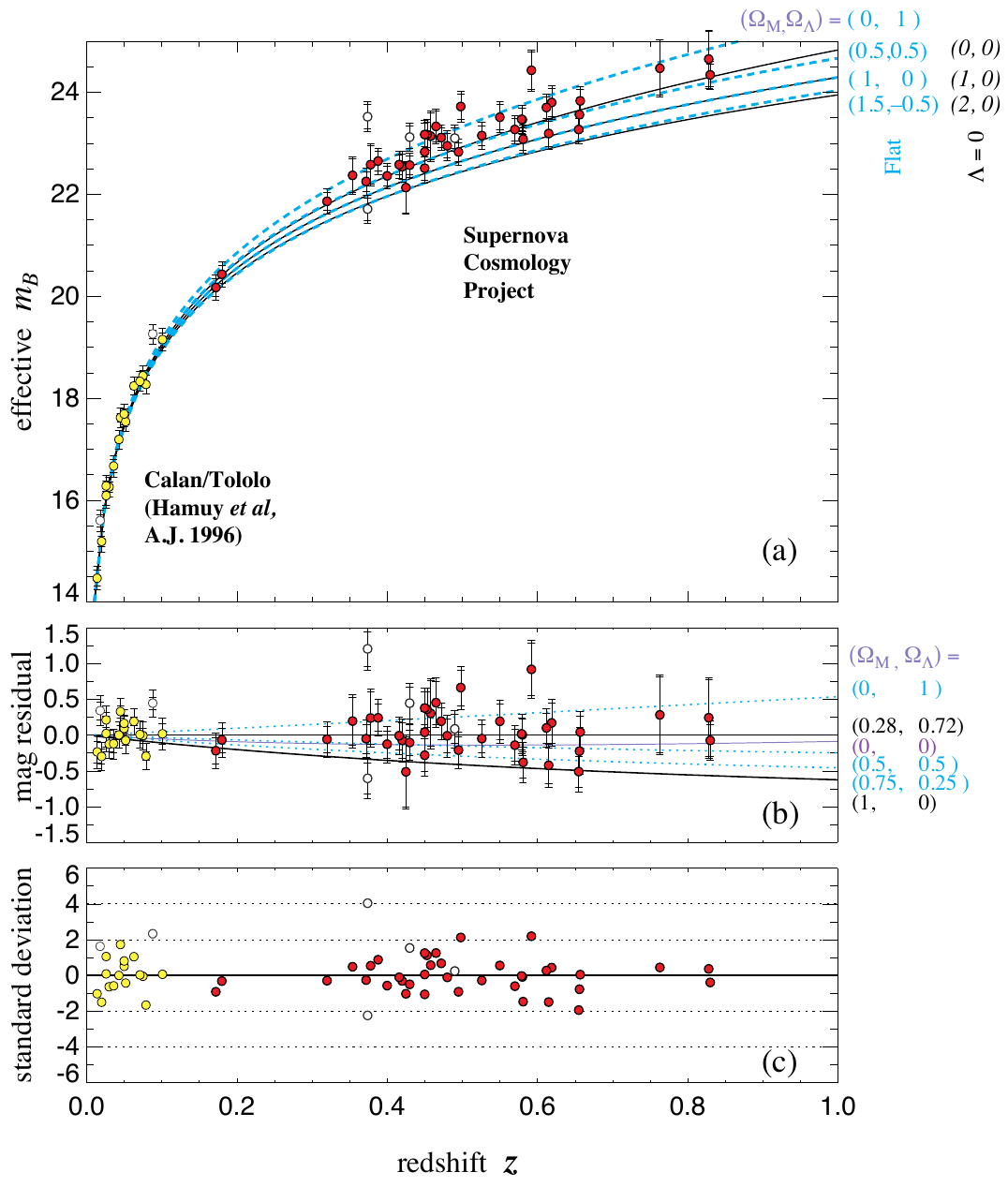}\qquad
    \includegraphics[width=.44\textwidth]{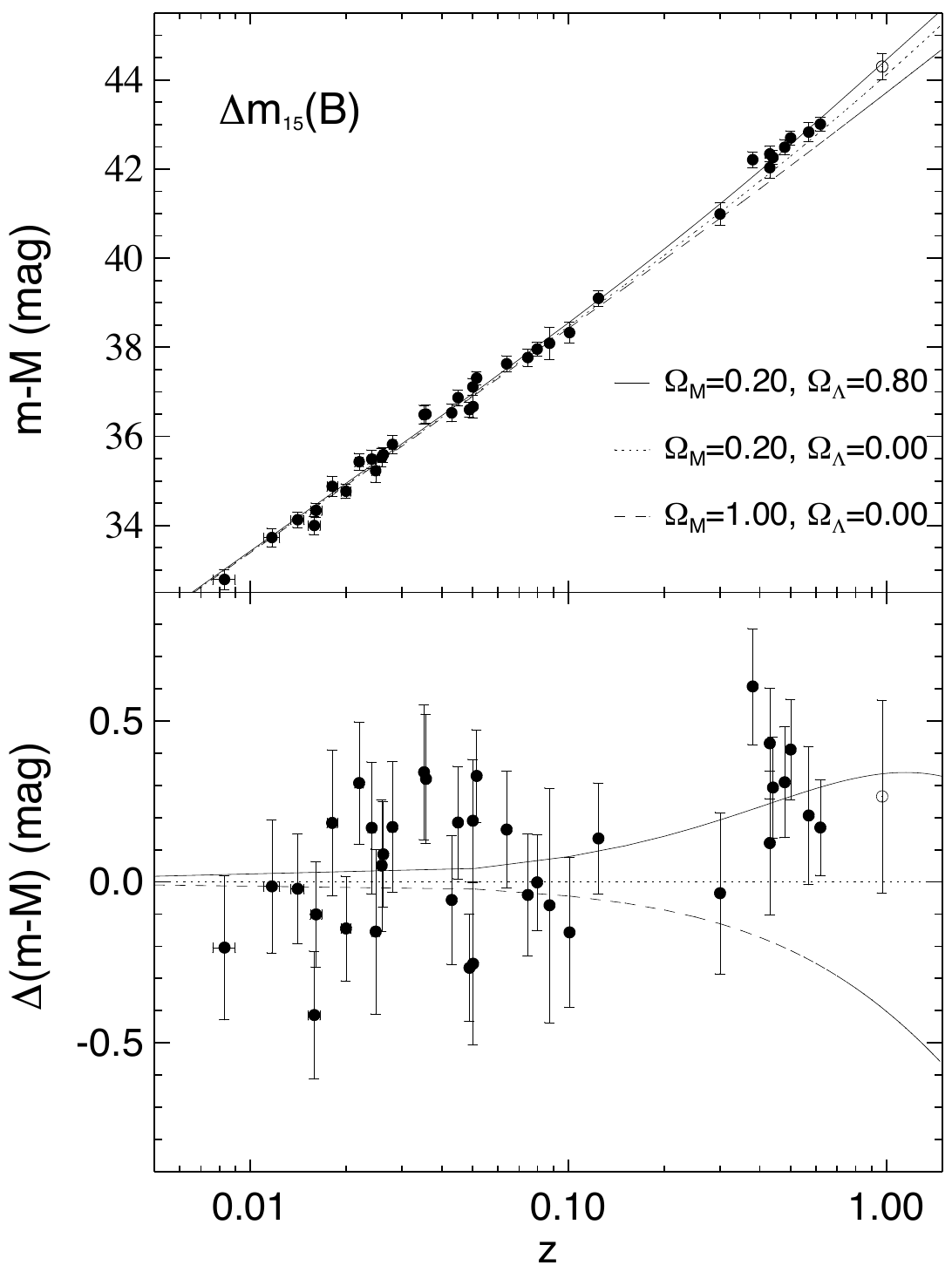}
    \caption{Observational data from SN Ia employed to infer that our Universe is experiencing accelerated expansion. \textbf{Left panel: }distance modulus data obtained by the Supernova Cosmology Project. \textbf{Right panel: }apparent magnitude data obtained by Supernova Search Team.\\
    \textit{Source:} \citet{Riess:1998cb,Perlmutter:1998np}.}
    \label{fig:sn}
    \end{figure}

    The cosmological constant has finally been incorporated into the dynamics of the Universe after consolidating the results obtained through the analysis of data from type Ia supernovae (SN Ia)\footnote{In fact, for some time there was an attempt to find some astrophysical argument to explain the results obtained with SN Ia.}. Using this methodology, two different groups, the Supernova Cosmology Project and the Supernova Search Team, found similar results, indicating that SN Ia appeared significantly fainter, as expected for a universe undergoing accelerated expansion~\citep{Riess:1998cb,Perlmutter:1998np}. The observational data employed by these two teams are shown in Fig.~\ref{fig:sn}. It is evident that there is a trend, particularly noticeable in the high redshift data, towards higher values of $\Omega_{\Lambda}$.

    Thus, with the conclusion of the accelerated expansion of the universe, however, without a complete understanding of its nature, the concept of DE was introduced in modern cosmology.

    \section{Theoretical aspects of the standard cosmological model}\label{sec:lcdm}
    
    This section is devoted to the presentation and discussion of the fundamental dynamics of the standard cosmological model, emphasizing the role of the cosmological constant in describing the observed accelerated expansion of the Universe. Additionally, we will address two key theoretical issues faced by the $\Lambda$CDM model, along with its observational aspects, achievements, and unresolved tensions.

    \subsection{The cosmological principle}\label{ssec:cosmprinc}
    
    On cosmological scales, systems are typically electrically neutral, so gravity is the dominant interaction. The standard cosmological model relies on Einstein's theory of General Relativity (GR) to describe gravity. Within GR, gravity manifests a geometrical aspect through distortions in space-time, which, differently from Newtonian mechanics, has a dynamical nature itself. In this context, the motion of free particles depends intrinsically on the geometry of spacetime. 
    
    The spacetime geometry is described by the metric tensor, which is a rank 2 covariant symmetric tensor. The metric tensor is a fundamental mathematical structure used to define distances and angles on differentiable manifolds, which serve as the natural framework for describing spacetime. In order to write the explicit expression for the metric that describes the Universe on cosmological scales, the Cosmological Principle (CP) is evoked. The CP postulates that matter in the Universe is distributed in a statistically homogeneous and isotropic manner on large scales. This means that any observer, regardless of its position in the Universe, is invariant under translations and rotations. Consequently, there are no preferred locations or directions, ensuring that observations made from any point in the universe (including Earth!) are representative. From a formal point of view, the CP implies that the 3D space is maximally symmetric. This condition leads to the well-known Friedmann-Lemaître metric, which, in polar coordinates, is given by
    \begin{equation}
        ds^{2}=g_{\mu\nu}dx^{\mu}dx^{\nu}=-dt^{2}+a\left(t\right)^{2} \left(\frac{dr^{2}}{1-Kr^{2}}+r^{2}d\theta^{2}+r^{2}\sin^{2}\left(\theta\right)d\phi^{2}\right)\,, \label{eq:metric}
    \end{equation}
    where $K$ is the space curvature\footnote{It is important to emphasize that, in order to satisfy the condition of maximal symmetry, the space curvature must be constant.} and $a\left(t\right)$ is a time dependent function called the scale factor. The scale factor is a crucial quantity to understand what the expansion of the Universe actually means. Within the domain of validity of the CP, it contains all the dynamics of spacetime. 
    
    As can be seen in Eq.~\eqref{eq:metric}, the scale factor is explicitly present in the line element of the 3D spatial foliation. Consequently, in the general case where it is time dependent, the distance between two stationary objects measured at different times will not coincide. In this context, it is important to understand that when we discuss the expansion of the Universe, objects themselves do not change their positions, but rather, space-time itself undergoes ``stretching'', a process described by the scale factor.

    \subsection{The composition of the Universe}\label{ssec:composition} 
    
    An important consequence of General Relativity is that the presence of matter can also deform the spacetime and consequently affect the motion of bodies. Therefore, it is of great importance to accurately describe the material composition of the Universe. 
    
    The CP allows us to describe the cosmic substratum as a fluid, and assuming the absence of heat flux or anisotropic stress in the universe, it can be described as a perfect fluid. Such a fluid is defined by an energy-momentum tensor of the following structure,
    \begin{equation}
        T^{\mu\nu}=\rho u^{\mu}u^{\nu}+p h^{\mu\nu} \,, \label{eq:tmunu}
    \end{equation}
    where $\rho$ is the energy density, $p$ is the pressure and $h^{\mu\nu}\equiv g^{\mu\nu}+u^{\mu}u^{\nu}$ is the so-called projection tensor.
    
    In the standard cosmological model, the material content of the Universe is composed of three components: radiation (relativistic matter, i.e. photons and relativistic neutrinos), baryons, and CDM. Radiation has dominated the universe's energy budget in the past, but today stands for merely a minimal fraction of it. Baryons are well-known ordinary matter, which can interact with radiation and constitute visible objects (like us!). Lastly, CDM is an exotic pressureless matter that seems to be incapable of significantly interacting with radiation and plays a crucial role in structure formation. It is further assumed that each of these individual components is also modeled as a perfect fluid with a constant equation of state $w_i=p_i/\rho_i$ where ($i={\rm r,b,c}$)\footnote{From now on the lower indices ${\rm r,b,c}$ denote radiation, baryons and CDM respectively. The absence of a lower index means the total cosmic medium.}. For the radiation component, the equation of state parameter is $w_{\rm r}=1/3$, and both baryons and CDM are pressureless, that is, $w_{\rm b}=w_{\rm c}=0$. Although baryons and CDM components are indistinguishable at the background, their distinction appears at the perturbative level, when it is considered the interaction between photons and baryons in the early universe, while CDM does not interact electromagnetically. In the background description, it is usual to combine these components into a single total matter component with $\rho_{\rm m}=\rho_{\rm b}+\rho_{\rm c}$ and $p_{\rm m}=p_{\rm b}=p_{\rm c}=0$.
    
    \subsection{The dynamics of the Universe in the $\Lambda$CDM model}\label{ssec:dyn}
    
    The standard cosmological model can be formulated from the following action,
    \begin{equation}
        \mathcal{S}=\int\left[\frac{1}{16\pi G}\left(R-2\Lambda\right)+\mathcal{L}_{\rm M}\right]\sqrt{-g}\ d^{4}x \,. \label{eq:action}
    \end{equation}
    The first term within the square brackets represents the standard Einstein-Hilbert action augmented by the cosmological constant, which describes the spacetime geometry. On the other hand, the second term within the square brackets denotes the Lagrangian for the entire cosmic medium, whose composition was detailed in Sect.~\ref{ssec:composition}. 
    
    Taking the variation of the action w.r.t. the metric, the well established Einstein field equations are obtained,
    \begin{equation}
        R_{\mu\nu}-\frac{1}{2}Rg_{\mu\nu}+\Lambda g_{\mu\nu}=8\pi G T_{\mu\nu}\,, \label{eq:einstein}
    \end{equation}
    where $R_{\mu\nu}$ is the Ricci tensor, and $R\equiv R_{\mu\nu}g^{\mu\nu}$ is its contraction, the so-called Ricci scalar. Whereas the left-hand side of Eq.~\eqref{eq:einstein} results from the variation of the geometric term in Eq.~\eqref{eq:action}, the right-hand side of Eq.~\eqref{eq:einstein} is obtained from the variation of the cosmic medium Lagrangian with respect to the metric, which results in the energy-momentum tensor given by Eq.~\eqref{eq:tmunu}.
    
    An important property of General Relativity is that the covariant derivative of the left-hand side of Eq.~\eqref{eq:einstein} must be identically zero. This condition is obtained from the Bianchi identities~\citep{bianchi1902sui}. As a trivial consequence, the covariant derivative of the right-hand side of Eq.~\eqref{eq:einstein} must also vanish, which can be interpreted as the conservation of the cosmic medium energy-momentum tensor. Since the universe is described as a perfect fluid, the spatial components of the covariant derivative of the energy-momentum tensor must be identically zero, while the time component delivers the conservation of total energy in the universe, expressed as,
    \begin{equation}
        \dot{\rho}+3H\left(\rho+p\right)=0 \,, \label{eq:continuity}
    \end{equation}
    where the dot denotes the derivative w.r.t. cosmic time, and $H\equiv\dot{a}/a$ is the Hubble parameter. 
    
    As a hypothesis of the $\Lambda$CDM model, each component of the universe is conserved independently, implying that there is no energy exchange between them. Consequently, an equation equivalent to Eq.~\eqref{eq:continuity} can be written for each individual component. By applying the respective equation of state to each component, one can solve the energy conservation equation to obtain the evolution of individual energy densities in terms of the scale factor,
    \begin{equation}
        \dot{\rho}_i+3H\rho_i\left(1+w_i\right)=0 \quad\Rightarrow\quad \rho_i=\rho_{i0}\ a^{-3\left(1+w_i\right)} \quad\Rightarrow\quad \left\{
    \begin{matrix} 
    \rho_{\rm r}=\rho_{\rm r0}\ a^{-4}\\ 
    \rho_{\rm m}=\rho_{\rm m0}\ a^{-3} 
    \end{matrix}
    \right. \,.  \label{eq:continuityi}
    \end{equation}
    The subscript 0 in a given quantity denotes its present value, with the standard convention of setting $a_0=1$ for normalization. According to Eq.~\eqref{eq:continuityi}, the total non-relativistic matter, comprising baryons and CDM, decays proportionally to $a^{-3}$, while radiation (relativistic matter) decays proportionally to $a^{-4}$. These expressions can be interpreted as follows: the density of non-relativistic matter decreases with the volume of the universe; on the other hand, relativistic matter not only decays with the volume of the universe, but also experiences a stretching of its wavelength due to cosmic expansion, resulting in an additional factor $a^{-1}$ in its evolution.
    
    At first sight, Einstein's equations seem to comprise 16 independent equations since they involve tensors of order 2. However, the symmetry of the metric reduces this number to 10, and the homogeneity and isotropy of the background level further reduce it to 2. The first independent equation can be derived from the 00 term of Eq.~\eqref{eq:einstein},
    \begin{equation}
        H^{2}=\frac{8\pi G}{3}\rho-\frac{K}{a^{2}}+\frac{\Lambda}{3}  \,.  \label{eq:friedmann}
    \end{equation}
    The Eq.~\eqref{eq:friedmann} is usually referred to as the Friedmann equation. The Friedmann equation establishes a connection between the dynamics of spacetime, encapsulated in the Hubble parameter\footnote{Note that describing spacetime dynamics in terms of either the scale factor or the Hubble parameter is equivalent, as the latter is derived from the former.}, and the matter distribution represented by $\rho$. These quantities are the degrees of freedom required for a complete description of the dynamics of the universe at the background level. It should be noted that, in this scenario, Einstein's equations form a closed system, as there are two degrees of freedom $H$ and $\rho$, and two independent equations. However, it is often convenient to replace the second independent Einstein equation with the conservation equation, which has already yielded individual solutions for the energy densities of the components of the universe. The total density of the universe is obtained by adding the individual densities. Substituting the solutions obtained in Eq.~\eqref{eq:continuityi} into Eq.~\eqref{eq:friedmann}, we derive the evolution of the Hubble parameter in terms of the scale factor. In general, there is no need to express the evolution of $H$ and $\rho$ in terms of cosmic time $t$, since the scale factor acts also as a parameterization of time, depicting the history of the universe from its beginning to the present within the range of zero to unity.
    
    Introducing the definitions of the density parameters,
    \begin{equation}
        \Omega_i\equiv\frac{8\pi G}{3H^{2}_0}\rho_{i0} \quad , \quad \Omega_{\rm k}\equiv-\frac{K}{H^{2}_0} \quad{\rm and}\quad \Omega_{\Lambda}\equiv\frac{\Lambda}{3H^{2}_0} \,,  \label{eq:omega}
    \end{equation}
    the solution for the Hubble parameter can be obtained directly from the Friedmann equation,
    \begin{equation}
        H=H_0\sqrt{\Omega_{\rm r}a^{-4}+\Omega_{\rm m}a^{-3}+\Omega_{\rm k}a^{-2}+\Omega_{\Lambda}}  \,.  \label{eq:hubble}
    \end{equation}
    While Eq.~\eqref{eq:hubble} seems to have five free parameters, it is important to note that as $a=a_0=1$, the Friedmann equation simplifies to a constraint between the cosmological parameters ($\Omega_{\rm r}+\Omega_{\rm m}+\Omega_{\rm k}+\Omega_{\Lambda}=1$), effectively reducing the number of free parameters to four. Based on this constraint, each density parameter signifies the fraction of its corresponding component in the present energy content of the universe. Ultimately, comparing observational data with theoretical quantities expressed in terms of these parameters allows us to find observational constraints to the values of these parameters and consequently to understand the current composition of the universe. 
    
    Even though the Friedmann equation and the continuity equation are sufficient for computing the background dynamics of the universe, Einstein's second independent equation sheds light on the necessity of the cosmological constant to account for the universe undergoing accelerated expansion. This equation arises from the combination of the Friedmann equation with the trace of Eq.~\eqref{eq:einstein}, resulting in the expression,
    \begin{equation}
        \frac{\ddot{a}}{a}=-\frac{4\pi G}{3}\left(\rho+3p\right)+\frac{\Lambda}{3}  \,. \label{eq:acceleration}
    \end{equation}
    As previously discussed, the dynamics of spacetime is contained within the scale factor (or equivalently, within the Hubble parameter). In an expanding universe, the scale factor describes the stretching of space, and an accelerated expansion indicates that the velocity of the stretching is increasing. Hence, the criterion for such acceleration is simply that the second derivative of the scale factor w.r.t. time is positive. However, according to Eq.~\eqref{eq:acceleration}, in the absence of a cosmological constant a positive $\ddot{a}$ is impossible since $\rho$ and $p$ are exclusively nonnegative. On the other hand, the presence of a positive cosmological constant opens the possibility of a positive second $\ddot{a}$. To ensure accelerated expansion, the condition to be satisfied by the cosmological constant is given by
    \begin{equation}
        \Lambda > 4\pi G\left(\rho+3p\right) \,.  \label{eq:condition}
    \end{equation}

    An important feature of the analysis presented here is that the cosmological constant can be represented as a fluid with a constant positive energy density and an equation of state $w_{\Lambda}=-1$, implying a negative pressure. Combining Eqs.~\eqref{eq:friedmann} and~\eqref{eq:acceleration}, it is straightforward to derive the energy density associated with the cosmological constant, expressed as follows,
    \begin{equation}
        \rho_{\Lambda}\equiv\frac{\Lambda}{8\pi G}\,,  \label{eq:condition}
    \end{equation}
    with $p_{\Lambda}=-\rho_{\Lambda}$.
    
\begin{figure}[t]
    \centering
    \includegraphics[width=.5\textwidth]{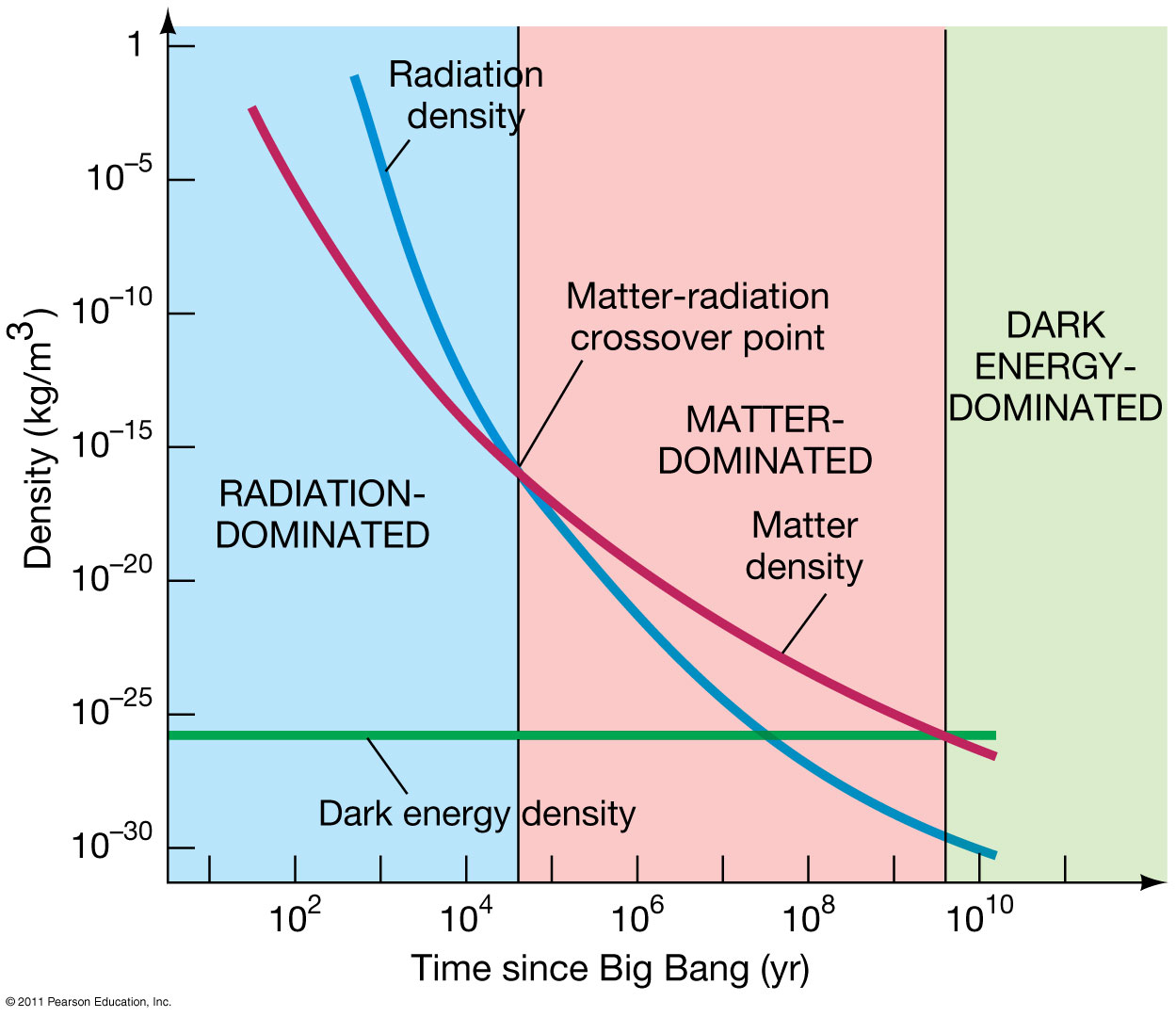}\qquad
    \caption{Time evolution of the energy density of the individual components of the Universe. \\
    \textit{Source:} \href{https://pages.uoregon.edu/jimbrau/BrauImNew/Chap27/7th/AT\_7e\_Figure\_27\_01.jpg}{https://pages.uoregon.edu/jimbrau/BrauImNew/Chap27/7th/AT\_7e\_Figure\_27\_01.jpg}.}
    \label{fig:density}
    \end{figure}

    By achieving a comprehensive understanding of the dynamics of the universe at the background level, we can visualize the time evolution of the energy density of its individual components. Using the results from Planck 2018, which suggest that the Universe is currently composed of approximately 30\% matter, 70\% DE, and a negligible fraction of radiation, the evolution of densities is depicted in Fig.~\ref{fig:density}. As illustrated, one can observe a phase of radiation domination in the past, succeeded by a phase of matter domination. Presently, our Universe is predominantly governed by DE.

    \subsection{Theoretical challenges of the $\Lambda$CDM model}\label{ssec:challenges}
    
    The formulation presented in Sect.~\ref{ssec:dyn}, while mathematically coherent, leaves some open questions. In general, these questions are connected to our ignorance about the nature of the constituents of the dark sector. In this section, two of these open questions are addressed: the cosmological constant problem; and the cosmic coincidence problem. There is hope that a better understanding of these issues may offer valuable insight into the nature of DE.
    
    \subsubsection{The cosmological constant problem}\label{ssec:ccp1}
    
    The most natural candidate for the cosmological constant is vacuum energy. As a consequence of the condition of being invariant under Lorentz transformations, the energy-momentum tensor of the vacuum must be proportional to the metric tensor, i.e., it can be written as
    \begin{equation}
        T_{\mu\nu}^{\rm (vac)}=-\frac{\Lambda^{\rm (vac)}}{8\pi G}g_{\mu\nu} \,.  \label{eq:tmunuvac}
    \end{equation}
    Upon comparing Eqs.~\eqref{eq:einstein} and~\eqref{eq:tmunuvac}, it is evident that the cosmological constant term appears in Einstein's equations in precisely the same form as the vacuum energy-momentum tensor, which makes this association immediate. On the other hand, quantum field theory predicts the vacuum energy density to be calculated as
    \begin{equation}
        \rho^{\rm (vac)}=\frac{\Lambda^{\rm (vac)}}{8\pi G}=\int_0^{k_{\rm lim}} \frac{d^{3}k}{\left(2\pi\right)^{3}}\frac{\sqrt{k^{2}+m^{2}}}{2}\approx\frac{k_{\rm lim}^{4}}{16\pi^{2}} \,,  \label{eq:rhovac}
    \end{equation}
    where $m$ and $k$ represent the mass and momentum of a scalar field used to calculate its zero-point energy (with $k_{\rm lim}\gg m$).
    
    The integral shown in Eq.~\eqref{eq:rhovac} diverges when k goes to infinity. Hence, a common practice is to introduce a cut-off scale as $k_{\rm lim}$, which defines the limit of validity of the theory. In the context of General Relativity, the Planck scale is a plausible candidate for this limit. However, employing this limit in Eq.~\eqref{eq:rhovac} yields a vacuum energy density of $10^{74}\ {\rm GeV}^{4}$, roughly 120 orders of magnitude larger than the observed results. This huge disparity has prevented the identification so far of the cosmological constant with vacuum energy. Nevertheless, advancements in our comprehension of quantum field theory within curved spacetime are expected to offer insights toward resolving this issue. A comprehensive discussion of the cosmological constant problem is available in the classical references~\citet{Weinberg:1988cp,Martin:2012bt}.

    \subsubsection{The cosmic coincidence problem}\label{ssec:ccp2}
    
    As discussed in Sect.~\ref{ssec:dyn}, a fundamental assumption of the standard cosmological model consists of considering that all material constituents of the universe are conserved independently. In other words, the evolution of the components in the universe are independent of each other. However, when analyzing the evolution of DE, it appears to have only become significant in recent epochs, as can be seen in Fig.~\ref{fig:density}. This observation raises an intriguing question: is it a coincidence that the component whose nature was more speculative was negligible in the past and became relevant recently? 
    
\begin{figure}[t]
    \centering
    \includegraphics[width=.317\textwidth]{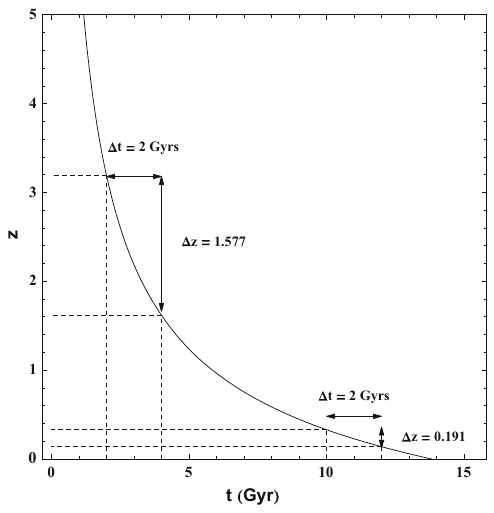}
    \includegraphics[width=.332\textwidth]{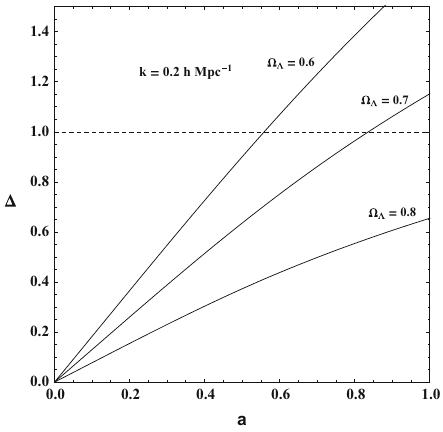}
    \includegraphics[width=.325\textwidth]{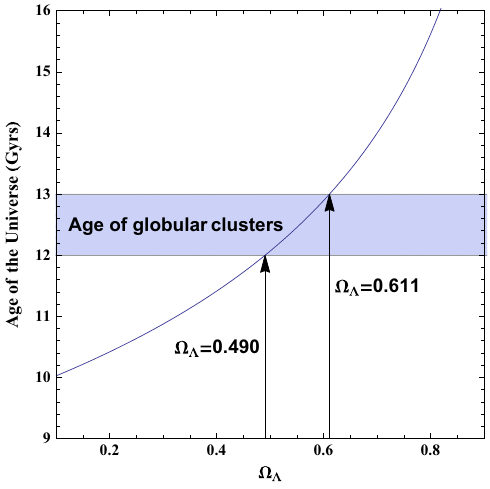}
    \caption{Three figures to illustrate the problem of cosmic coincidence. \textbf{Left panel: }redshift in the Universe in terms of cosmic time. \textbf{Center panel: }evolution density contrast for different values of $\Omega_{\Lambda}$ on typical scales of galaxy clusters. \textbf{Right panel: }age of the Universe in terms of $\Omega_{\Lambda}$. \\
    \textit{Source:} \citet{Velten:2014nra}.}
    \label{fig:coincidence}
    \end{figure}

    Analyzing the time evolution in terms of the redshift\footnote{This choice is convenient in most of the cases because, besides of serving as a parameterization for time (as well as the scale factor), redshift can be directly inferred by observations. In this case, the beginning of the universe is characterized at $z\rightarrow\infty$, and the present epoch is $z=0$.}, defined as $z=1/a-1$, the DE overcomes matter in $z\approx0.3$, which seems to be ``almost now''. In reality, the statement of the cosmic coincidence problem claiming that DE becomes significant only ``today'' sounds inaccurate. DE dominance has been lasting for roughly one-third of the history of the Universe, which is more than the radiation-dominated era. This apparent short time is because the relationship between redshift and cosmic time is non-linear. In other words, redshift intervals at different epochs do not represent the same time interval, as depicted in the left panel of Fig.~\ref{fig:coincidence}

    However, this does not imply that the coincidence problem is not a real problem. Actually, it can be reformulated on the basis of the viability of the formation of large-scale structures in the Universe and the age of the oldest astrophysical objects. Regarding the formation of large-scale structures, a key quantity for describing their evolution is the matter density contrast, defined as $\Delta\equiv\delta\rho_m/\rho_m$, where $\delta\rho_m$ denotes a perturbation of the matter energy density. Intuitively, $\Delta$ quantifies the size of the perturbations compared to the background. At the linear level, the matter density contrast obeys the Jeans equation,
    \begin{equation}
        \ddot{\Delta}+2H\dot{\Delta}+\left(\frac{c_s^{2}k^{2}}{a^{2}}-4\pi G\rho\right)\Delta=0 \,,  \label{eq:jeans}
    \end{equation}
    where $c_s^{2}$ is the square of the speed of sound of the fluid and $k$ is the comoving wavenumber.
    
    For structures to actually form, they must reach the non-linear level. A good estimate to know whether structures of a given scale reach the non-linear level is to solve the Jeans equation to obtain when the density contrast reaches values close to unity. The center panel of Fig.~\ref{fig:coincidence} shows the solutions of the matter density contrast for different values of $\Omega_{\Lambda}$ considering typical scales of galaxy clusters ($k=0.2\ h{\rm Mpc}^{-1}$). As can be seen, if $\Omega_{\Lambda}$ is much bigger than 70\%, the formation of galaxy clusters is impossible. 
    
    Moreover, it is obvious that the universe must be older than its oldest objects. Hence, knowing the ages of various objects enables one to estimate the minimum age of the universe. Theoretically, the age of the universe can be computed as,
    \begin{equation}
        t_{0}=\int_{0}^{t_0}dt=\int_{0}^{t_0}\frac{da}{aH\left(a\right)} \,,  \label{eq:age}
    \end{equation}
    where, within the $\Lambda$CDM model framework, the Hubble parameter is given by Eq.~\eqref{eq:hubble}.
    
    Objects called globular clusters are among the oldest systems we know in the universe, with an age of around 12-13 Gyrs. Note that these values exclude the possibility of a universe composed only of matter, whose age would be about 9 Gyrs. Increasing the contribution of the cosmological constant increases the age of the Universe. The right panel of Fig.~\ref{fig:coincidence} shows the age of the Universe as a function of $\Omega_{\Lambda}$. In the same figure, the typical age of the globular clusters is highlighted, indicating that, to satisfy the condition of having globular clusters, it must have $\Omega_{\Lambda}>0.6$. Considering also that there must be some incubation time for the formation of these globular clusters, we come close to the value obtained observationally.
    
    Thus, although the narrative that DE has only become relevant today in the dynamics of the universe might be considered to be not accurate, it seems that there is a coincidence in the fact that we live in a universe that has exactly the amount of DE necessary to explain the formation of structures and age of objects in the way we observe.

    \subsection{Observational aspects of the $\Lambda$CDM model}\label{ssec:observ}
    
    As in any area of Physics, experimental verification of a given hypothesis/theory is a vital process in cosmology. This section is devoted to discuss some of the main observational aspects of the $\Lambda$CDM model, including the most up-to-date results and the recent tensions with the results from Planck 2018.
    
    \subsubsection{SN Ia analysis}\label{ssec:snia}
    
    Given its historical importance as the first direct observation of the accelerated expansion of the universe, a particular emphasis will be given on the formalism employed to compare observational data on SN IA with theory. 
    
    A SN Ia emerges as the final stage of the accretion/merger process involving a carbon-oxygen white dwarf within a binary system. When the white dwarf reaches the Chandrasekhar mass, given by $1.4\ M_{\odot}$, the gravitational force overcomes the pressure of the degenerate electron gas~\citep{shapiro1983physics}. As a result, the system collapses giving rise to a huge explosion, as bright as a galaxy. According to this astrophysical mechanism, the SN Ia has a characteristic (constant) absolute magnitude, which makes the SN Ia suitable to be considered as ``standardizable'' candles. Given that the absolute magnitude can be calibrated for all SN Ia, it becomes feasible to obtain their distance based on the difference between their absolute and apparent magnitudes, defined as the distance modulus. The distance modulus quantifies the magnitude difference caused by light propagation through the universe, including curvature and expansion effects. The theoretical expression for the distance modulus is
    \begin{equation}
        \mu=m-M=5\log\left(\frac{d_{L}}{\rm 1Mpc}\right)+25 \,,  \label{eq:mu}
    \end{equation}
    where, $m$ and $M$ are the apparent and absolute magnitudes respectively, and $d_{L}$ is the luminosity distance, which is given by,
    \begin{equation}
    d_{L}=\left\{
    \begin{matrix*}[l]
      \dfrac{1}{\left(1+z\right)H_{0}\sqrt{|\Omega_{k}|}}\sin\left[\sqrt{|\Omega_{\rm k}|}H_0\stretchint{5ex}_{0}^{z}\dfrac{d\tilde{z}}{H\left(\tilde{z}\right)}\right] & {\rm for}\ \Omega_k>0 \,, \vspace{2mm} \\
      \dfrac{1}{\left(1+z\right)}\stretchint{5ex}_{0}^{z}\dfrac{d\tilde{z}}{H\left(\tilde{z}\right)} & {\rm for}\ \Omega_k=0 \,, \vspace{2mm} \\ 
      \dfrac{1}{\left(1+z\right)H_{0}\sqrt{|\Omega_{k}|}}\sinh\left[\sqrt{|\Omega_{\rm k}|}H_0\stretchint{5ex}_{0}^{z}\dfrac{d\tilde{z}}{H\left(\tilde{z}\right)}\right] & {\rm for}\ \Omega_k<0 \,, \vspace{2mm} \,.
    \end{matrix*}\right.
    \end{equation}

    The SN Ia surveys measure their apparent magnitudes, and when performing the statistical analysis to constrain the cosmological parameters, the absolute magnitude is accounted as a nuisance parameter. As an example, Fig.~\ref{fig:sn} displays the observational data used to first demonstrate the accelerated expansion of the universe. 
    
\begin{figure}[t]
    \centering
    \includegraphics[width=.655\textwidth]{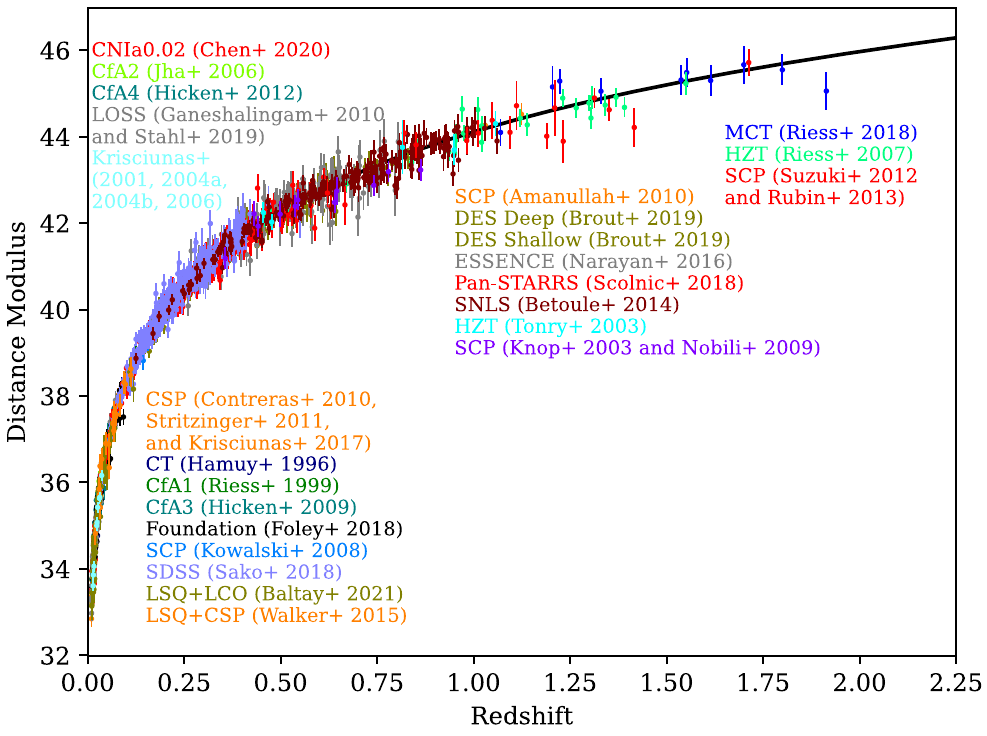}
    \includegraphics[width=.29\textwidth]{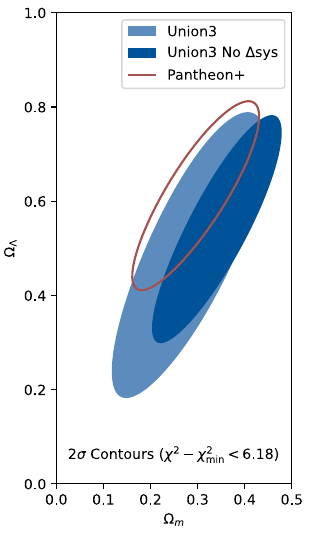}
    \caption{Current data and results from SN Ia data. \textbf{Left panel: }Union 3 distance modulus data, consisting of a compilation of various observations represented by different colors in the figure. The solid black line consists to the $\Lambda$CDM best fit. \textbf{Center panel: }contour plot on the $\Omega_{m}-\Omega_{\Lambda}$ plane. The blue shaded regions represent the results obtained from Union 3 data, while the purple line corresponds to the same results obtained from the Pantheon+ SN Ia sample. \\
    \textit{Source:} \citet{Rubin:2023ovl}.}
    \label{fig:snmodern}
    \end{figure}

    Currently, the largest available SN Ia dataset is Union 3~\citep{Rubin:2023ovl}, which contains 2,087 cataloged objects. These data, along with the Union 3 constraints for $\Omega_{m}$ and $\Omega_{\Lambda}$ are presented in Fig.~\ref{fig:snmodern}. The impressive increase in the number and quality of the SN Ia data can be visualized by comparing the left panel with Fig.~\ref{fig:sn}. In the right panel, the $\Omega_{m}-\Omega_{\Lambda}$ plane obtained in the Union 3 statistical analysis is shown. This result corroborates with an accelerated expansion, but delivers slightly more matter than usual (still within $2\sigma$ agreement with Planck 2018). It should be mentioned that Union 3 is a compilation of SN Ia measurements accumulated over time by several surveys. Recently, the DES Collaboration conducted a cosmological analysis using its own catalog of 1635 SN Ia~\citep{DES:2024tys}. The results obtained by DES are in accordance with the results of Union 3.

    \subsubsection{Other cosmological probes}\label{ssec:probes}
    
    Beyond SN Ia, the $\Lambda$CDM model has undergone recurrent testing using a variety of cosmological data. The diversity of observational tests is crucial because it mitigates the risk of erroneous conclusions based on biases induced by observational systematic errors, which are unique to each experiment. Among others, two observational probes of great importance in modern cosmology are:
    \begin{itemize}
        \item \textbf{Cosmic Microwave Background (CMB):} The CMB is a relic radiation from the early universe, approximately 380,000 years post the initial singularity, when the coupling between baryons and photons ceased at the last scattering surface. This radiation is now observed across the sky as a background radiation with a temperature of about 2.73 K. The analysis of temperature anisotropies, polarization, and lensing effects in the CMB map yields robust constraints on the cosmological parameters of the $\Lambda$CDM model. The most recent results on CMB data have been provided by Planck 2018~\citep{Planck:2018vyg}. \vspace{2mm}
    
        \item \textbf{Baryon Acoustic Oscillations (BAO):} The BAO consists of a regular spherical pattern imprinted in the density of baryonic matter, appearing as a small overdensity. These oscillations originate from the freezing of propagating sound waves in the early photon-baryon fluid. As the universe evolves, the formation of cosmic structures is favored in these regions, allowing their detection through the distribution of galaxies. BAO measurements have played a crucial role in constraining cosmological parameters, particularly in determining the curvature of the Universe. The main results regarding the BAO data have been provided by the Sloan Digital Sky Survey (SDSS)~\citep{Beutler:2012px,Ross:2014qpa,BOSS:2016wmc,eBOSS:2020lta,eBOSS:2020abk,eBOSS:2020gbb,eBOSS:2020tmo}. \vspace{2mm}
    \end{itemize}
    
\begin{figure}[t]
    \centering
    \includegraphics[width=.4\textwidth]{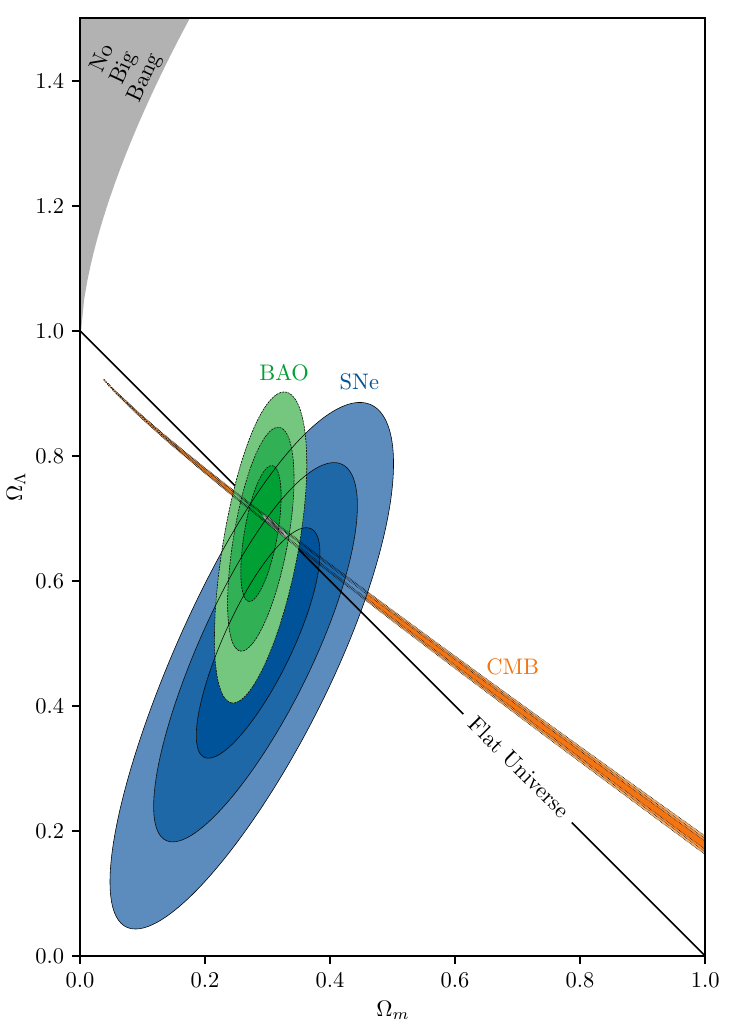}
    \caption{Most updated observational constraints for the LCDM model. The blue, green, and orange regions show the contours for the SN Ia, BAO, and CMB analyzes, respectively. \\
    \textit{Source:} \citet{Rubin:2023ovl}.}
    \label{fig:contour}
    \end{figure}

    The most impressive achievement of the standard cosmological model is to be able to precisely fit the data obtained from various cosmological probes using consistent values for the cosmological parameters. This consistency of parameters in the observational results is exactly what is expected from the ``correct'' description of the universe, which gave the $\Lambda$CDM model the status of a concordance model. The probability contours in the $\Omega_{m}-\Omega_{\Lambda}$ plane obtained from the SN Ia, BAO, and CMB analyzes are shown in Fig.~\ref{fig:contour}, where it is possible to see that there is an intersection region that characterizes the concordance between the probes. The values obtained in these analyzes are presented in Tab.~\ref{tab:resultlcdm}.
    \begin{table}[]
    \centering
    \begin{tabular}{lccc}
    \hline\hline
    \multicolumn{4}{c}{\multirow{2}{*}{Flat $\Lambda$CDM}}                                                             \\
    \multicolumn{4}{c}{}                                                                                               \\ \hline
    Probe             & $H_0^{\dagger}$   & $\Omega_{\rm m}$          & $\Omega_{\rm k}$          \\ \hline
    SN Ia             & -               & $0.356^{+0.028}_{-0.026}$ & -                         \\
    BAO + CMB         & $67.7\pm0.4$    & $0.311\pm0.006$           & -                         \\
    SN Ia + BAO + CMB & $67.5\pm0.4$    & $0.313\pm0.006$           & -                         \\ \hline
    \multicolumn{4}{c}{\multirow{2}{*}{Non-flat $\Lambda$CDM}}                                                         \\
    \multicolumn{4}{c}{}                                                                                               \\ \hline
    SN Ia             & -                                      & $0.287^{+0.064}_{-0.066}$ & $0.203^{+0.183}_{-0.173}$ \\
    BAO + CMB         & $67.9\pm0.6$                           & $0.310\pm0.006$           & $0.001\pm0.002$           \\
    SN Ia + BAO + CMB & $67.7\pm0.6$                           & $0.312\pm0.006$           & $0.001\pm0.002$           \\ \hline\hline
    \end{tabular}
    \caption{Observational results of the $\Lambda$CDM model.\\ 
    $^{\dagger}$Due to a degeneracy between $H_0$ and the SN Ia absolute magnitude, the analysis with only SN Ia data can not constraint $H_0$. \\
    \textit{Source:} Adapted from Tab. 6 of~\citet{Rubin:2023ovl}.}
    \label{tab:resultlcdm}
    \end{table}

    \subsubsection{Cosmological tensions}\label{ssec:tension}
    
    Despite of the observational success of the $\Lambda$CDM model, some recent results have found notable discrepancies compared to the results derived from Planck 2018 data. The two most prominent cases occur for: ($i$) weak lensing measurements regarding the amplitude of matter perturbations, encoded in the parameter $\sigma_8$, which quantifies the root-mean-square of matter density fluctuations on a scale of $8\,h^{-1}{\rm Mpc}$\footnote{In reality, this discrepancy is better quantified in terms of the parameter $S_8=\sqrt{\Omega_{\rm m}/0.3}\,\sigma_8$.}; and ($ii$) local measurements of the Hubble constant $H_0$, which are typically obtained using the distance ladder method, linking the distances to nearby astronomical objects through standard candles. These issues are commonly referred to in the recent literature as $S_8$ tension and $H_0$ tension, respectively. An argument suggests that this tension arises between early and late-time observations~\citep{Verde:2019ivm} (cf.~\citet{Linder:2021ujs}).
    
    Regarding structure formation, recent weak lensing results from the Kilo Degree Survey (KiDS) indicate approximately a $3\sigma$ tension in the amplitude of matter scalar perturbations, characterized by the $S_8$ parameter, compared to the latest Planck 2018 results~\citep{Heymans:2020gsg,KiDS:2020suj}. It should be noted that a recent similar analysis from the Dark Energy Survey (DES) suggests a difference of around $2\sigma$ from the Planck 2018 results, which is considered acceptable support for the $\Lambda$CDM model~\citep{DES:2021wwk}. A more substantial disparity arises in the determination of the current value of the Hubble constant. Although early-time observations from Planck 2018 and DES+BAO+BBN yield $H_{0}=67.66\pm0.42 \text{ km s}^{-1} {\rm Mpc}^{-1}$ and $H_{0}=67.6\pm0.9 \text{ km s}^{-1} {\rm Mpc}^{-1}$, respectively, late-time local measurements from SH0ES analysis with Cepheid variables yield $H_{0}=74.04\pm1.04 \text{ km s}^{-1} {\rm Mpc}^{-1}$~\citep{Riess:2021jrx}. This inconsistency currently exceeds the $4\sigma$ confidence level. A recent similar analysis combining data from Cepheids, Tip of the Red Giant Branch (TRGB) and Surface Brightness Fluctuations (SBF) delivered $H_{0}=71.76\pm0.58\ ({\rm stat})\ \pm1.19\ ({\rm sys}) \text{ km s}^{-1} {\rm Mpc}^{-1}$~\citep{Uddin:2023iob}, which is more in agreement with Planck 2018. 
    
\begin{figure}[t]
    \centering
    \includegraphics[width=.395\textwidth]{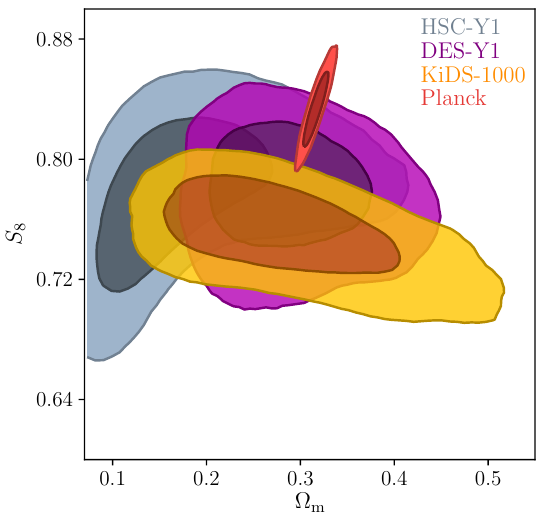}
    \includegraphics[width=.38\textwidth]{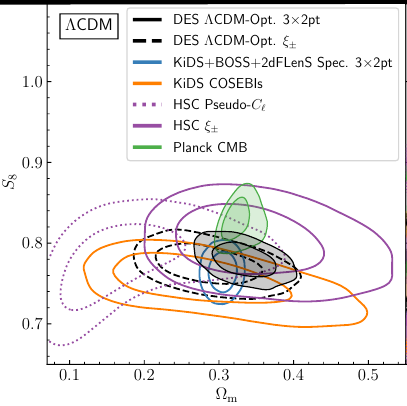}
    \includegraphics[width=.8\textwidth]{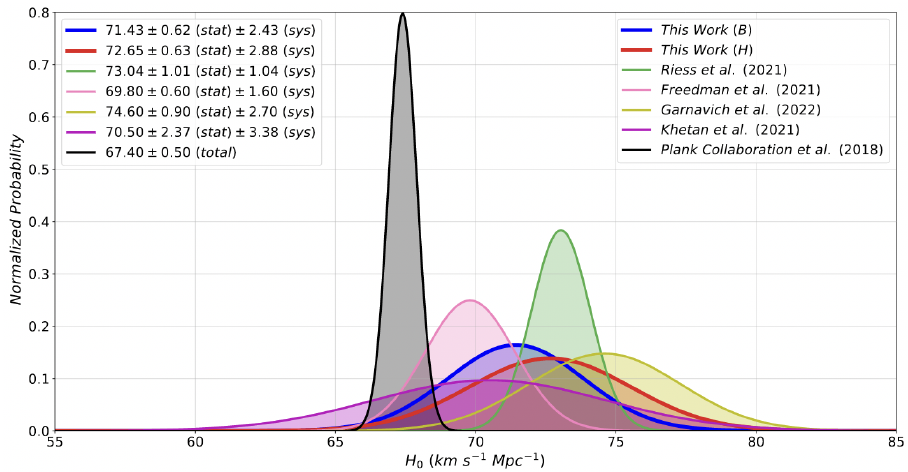}
    \caption{Recent observational results that support the $S_8$ tension and $H_0$ tension. \textbf{Upper left panel: }KiDS constraint for the $\Omega_{\rm m}-S_8$ plane. \textbf{Upper right panel: }DES constraint for the $\Omega_{\rm m}-S_8$ plane. \textbf{Bottom panel: }posteriors for $H_0$ from Planck 2018 and local measurements. \\
    \textit{Source:} \citet{KiDS:2020suj,DES:2021wwk,Freedman:2023jcz}.}
    \label{fig:tension}
    \end{figure}

    The key observational results that motivate $S_8$ tension and the $H_0$ tension are shown in Fig.~\ref{fig:tension}. The upper left and upper right panels show the results for $S_8$ from KiDS and DES, respectively, where one can see that the overlap between the contours from Planck 2018 and KiDS in the left panel is slightly smaller than between Planck 2018 and DES (black and green regions). On the other hand, the bottom panel shows the posteriors for $H_0$ from Planck 2018 and several local measurements. Concerning the $H_0$ results, it is important to emphasize that the Planck 2018 prediction is based on the $\Lambda$CDM, while the results from local measurements are model-independent. This is a robust argument in favor of new physics. A more complete description of the current status of the tension $H_{0}$ can be found in Refs.~\citet{Riess:2019qba,Freedman:2023jcz}.

    \section{DE models}\label{sec:models}
    
    Within the context of limited understanding of the nature of the cosmological constant and recent observations deviating from $\Lambda$CDM predictions, numerous alternative theories have emerged and will continue to be proposed in the attempt to explain the acceleration of the Universe. This section is dedicated to exploring the main alternatives related to the presence of an additional exotic DE component.
    
    \subsection{Quintessence models}\label{ssec:scf}
    
    The simplest approach to introduce an unknown component that permeates spacetime is to incorporate a scalar field with a canonical kinetic term and under the action of a given potential. In cosmological terms, this scalar field is called quintessence. The dynamics of the universe within this framework can be formulated through the following action
    \begin{equation}
        \mathcal{S}=\int\left[\frac{1}{16\pi G}R+\mathcal{L}_{\phi}+\mathcal{L}_{\rm M}\right]\sqrt{-g}\ d^{4}x \,, \label{eq:actionscf}
    \end{equation}
    where $\mathcal{L}_{\rm M}$ is the lagrangian of the ordinary cosmic medium, composed by radiation, baryons and CDM, and $\mathcal{L}_{\phi}$ is the lagrangian of the scalar field. Note that the cosmological constant is not considered in this framework. More precisely, the scalar field lagrangian is given by, 
    \begin{equation}
        \mathcal{L}_{\phi}=-\frac{1}{2}g^{\mu\nu}\partial_{\mu}\phi\partial_{\nu}\phi-V\left(\phi\right) \,, \label{eq:lagrangianscf}
    \end{equation}
    where the first term in the Lagrangian represents the canonical kinetic term for the scalar field, while the second term denotes the potential, which is a function of the scalar field.
    
    In this case, there are now two dynamical fields: the metric tensor and the scalar field. Thus, the dynamics of the universe is obtained by taking the variation w.r.t. them. Taking the variation w.r.t. the metric delivers the Einstein equations in the context of quintessence,
    \begin{eqnarray}
        H^{2}&=&\frac{8\pi G}{3}\left[\frac{\dot{\phi}^{2}}{2}+V\left(\phi\right)+\rho\right] \,, \label{eq:friedmannscf} \\
        \dot{H}&=&-\frac{4\pi G}{3}\left(\dot{\phi}^{2}+\rho+p\right) \,. \label{eq:accelerationscf}
    \end{eqnarray}

    On the other hand, taking the variation w.r.t. the scalar field gives the Klein-Gordon equation,
    \begin{equation}
        \ddot{\phi}+3H\dot{\phi}+V_{,\phi}=0 \,, \label{eq:kgscf} 
    \end{equation}
    where $V_{,\phi}\equiv dV/d\phi$\footnote{From now on the dependency of the scalar field in the potential will be omitted.}. 
    
    In order to obtain explicitly the dynamics of the Universe, a particular expression for the potential must be chosen. Each quintessence model is characterized by the choice of the potential. As is done for theories of inflation, the ability of a potential to generate an accelerated expansion can be quantified by the following slow-roll parameters,
    \begin{equation}
        \epsilon\equiv\frac{1}{16\pi G}\left(\frac{V_{,\phi}}{V}\right)^{2}\qquad{\rm and}\qquad\eta\equiv\frac{1}{4\pi G}\frac{V_{,\phi\phi}}{V} \,. \label{eq:slowroll} 
    \end{equation}
    The closer $\epsilon$ and $\eta$ are to zero, the faster is the expansion of the universe. In the limiting case $\epsilon=0$ and $\eta=0$ the expansion is exponential.  
    
    A typical discrimination between quintessence models is done by dividing the models into two groups: freezing models~\citep{Ferreira:1997hj,Caldwell:1997ii,Carroll:1998zi} and thawing models~\citep{delaMacorra:1999ff,Kallosh:2003bq}. In freezing models the field is free to roll in the past, however, in recent times the field reaches the slow-roll regime, which leads to the accelerated expansion. On the other hand, in thawing models the field is frozen in the past but starts to evolve in recent times.  
    
    One important feature of the scalar field approach is that, similarly to the fact that the cosmological constant can be rewritten as a fluid with constant energy density and $w_{\Lambda}=-1$, it is also possible to define a fluid that will provide the same dynamics as the scalar field. More precisely, one can define an energy density and pressure in terms of the scalar field and the potential. An important difference from the cosmological constant case is that, since the scalar field is dynamical, the resulting fluid is more general in the sense that it might be time dependent. Using Eqs.~\eqref{eq:friedmannscf} and~\eqref{eq:kgscf} it is easy to conclude that the fluid description associated to the scalar field is given by,
    \begin{equation}
    \left\{
    \begin{matrix*}[l]
      \rho_{\phi}&=& \dfrac{1}{2}\dot{\phi}^{2}+V\left(\phi\right) \vspace{2mm} \\
      p_{\phi}&=& \dfrac{1}{2}\dot{\phi}^{2}-V\left(\phi\right) 
    \end{matrix*}
    \right. \qquad\Rightarrow\qquad w_{\phi}=\frac{\dot{\phi}^{2}+2V\left(\phi\right)}{\dot{\phi}^{2}-2V\left(\phi\right)}\,. \label{eq:fluidscf} 
    \end{equation}
    The slow-roll conditions is reached when $\dot{\phi}^{2}\rightarrow0$, and consequently $w_{\phi}\rightarrow-1$.
    
    An important feature of the quintessence models is that the canonical kinetic term is associated with a luminal sound speed, which means that the effect of pressure perturbations overcomes the gravitational colapse at the perturbative level. As a consequence, quintessence DE is smooth like the cosmological constant. i.e., it does not cluster.
    
    \subsection{$k$-essence models}\label{ssec:dyn}
    
    The quintessence models can be generalized to the case where the kinetic term is no longer canonical, but a function of the field and its kinetic energy $X=-\left(1/2\right)g^{\mu\nu}\partial_{\mu}\phi\partial_{\nu}\phi$. This kind of model is named as $k$-essence model~\citep{Armendariz-Picon:2000ulo}. In this case, the action presented in Eq.~\eqref{eq:actionscf} is generalized as
    \begin{equation}
        \mathcal{S}=\int\left[\frac{1}{16\pi G}R+\mathcal{F}\left(X,\phi\right)+\mathcal{L}_{\rm M}\right]\sqrt{-g}\ d^{4}x \,. \label{eq:actionk}
    \end{equation}
    Specific models are characterized by the explicit choice of $\mathcal{F}\left(X,\phi\right)$. Some examples of $k$-essence models can be found in~\citep{Silverstein:2003hf,Garousi:2000tr,Alishahiha:2004eh}. Although $k$-essence models are more general than quintessence models, it is also possible to write an equivalent fluid. In other words, there are an energy density and pressure associated with the dynamics of the $k$-essence scalar field,
    \begin{equation}
    \left\{
    \begin{matrix*}[l]
      \rho_{\phi}&=& 2X\mathcal{F}_{,X}-\mathcal{F} \vspace{2mm} \\
      p_{\phi}&=& \mathcal{F} 
    \end{matrix*}
    \right. \qquad\Rightarrow\qquad w_{\phi}=\frac{\mathcal{F}}{2X\mathcal{F}_{,X}-\mathcal{F}}\,. \label{eq:fluidscf} 
    \end{equation}
    In the slow-roll regime $2X\mathcal{F}_{,X}$ tends to zero, and consequently $w_{\phi}\rightarrow-1$.
    
    An important difference in comparison with quintessence is that, due to the fact that the kinetic term is no longer the canonical one, the sound speed is not equal to the speed of light. In this case, the sound speed must be taken into account in the perturbation equations, and DE might cluster. The $k$-essence sound speed can be obtained doing $c_s^{2}=p_{\phi,X}/\rho_{\phi,X}$, which results in,
    \begin{equation}
        c_s^{2}=\frac{\mathcal{F}_{,X}}{\mathcal{F}_{,X}+2X\mathcal{F}_{,XX}} \,. \label{eq:csk}
    \end{equation}
    In fact, the sound speed is a key quantity related to the stability of the growing structures. A classical condition for the stability of the growing modes is characterized by a positive sound speed. Nonetheless, in order to also satisfy some criteria for quantum stabilities, both the numerator and the denominator of Eq.~\eqref{eq:csk} must be positive.
    
    \subsection{Fluid approach}\label{ssec:fluid}
    
    Since both quintessence and $k$-essence models can be mapped into fluids, instead of choosing a given potential $V\left(\phi\right)$ or a function $\mathcal{F}\left(X,\phi\right)$, a common approach is to propose a general equation of state for the DE fluid. To illustrate the equivalence between the approaches, let us assume a given quintessence model. Assuming a given proposed potential, the Klein-Gordon equation, Eq.~\eqref{eq:kgscf}, can be used to compute the scalar field solution and, subsequently, Eq.~\eqref{eq:fluidscf} determines the equation of state. 
    
    The two most explored descriptions in the literature are:
    \begin{itemize}
        \item \textbf{$w$CDM model:} In this first case the equation of state of DE is still constant, but is not necessarily -1. The equation of state of the DE fluid can be written as $p_{x}=w_{0}\rho_{x}$\footnote{The DE fluid will be denoted with a lower index $x$.}, where $w_{0}$ is considered a free parameter. This approach can be interpreted as a first extension of the cosmological constant case.
        \item \textbf{CPL parameterization:} In this model it is proposed a linear parameterization for the DE equation of state, i.e., the explicit expression for the DE equation of state is,
        \begin{equation}
            w_x=w_0+\left(1-a\right)w_a \,. \label{eq:wxcpl}
        \end{equation}
        In fact, this approach is more complex. First, the DE fluid is described by two extra parameters. Furthermore, the DE equation of state is now time-dependent. Regarding the evolution of $w_x$, Eq.~\eqref{eq:wxcpl} can be interpreted as follows: $w_0$ represents the final value reached in current times, and $w_a$ modulates the time evolution in the past. 
    \end{itemize} \vspace{2mm}
    Other examples of parameterizations for DE can be found in~\citet{Wetterich:2004pv,Barboza:2008rh}.
    
\begin{figure}[t]
    \centering
    \includegraphics[width=.45\textwidth]{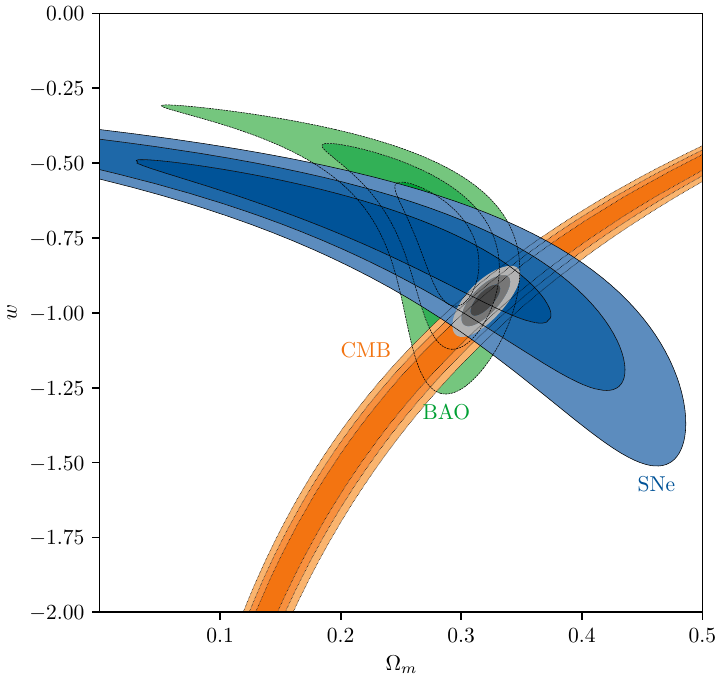}
    \includegraphics[width=.505\textwidth]{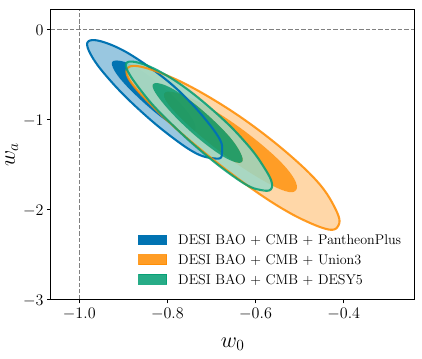}
    \caption{Recent observational constraints for $w$CDM and CPL models. \textbf{Left panel: }Results from Union 3 in the $w$CDM model. The blue, green and orange regions show the contours for the SN Ia, BAO, and CMB analyzes, respectively. \textbf{Right panel: }DESI results in the CPL model.  The blue, green and orange regions show the contours of DESI+CMB combined with different SN Ia surveys. \\
    \textit{Source:} \citet{Rubin:2023ovl,DESI:2024mwx}.}
    \label{fig:resultwcdm}
    \end{figure}

    Since both $w$CDM and CPL can be seen as simple effective extensions of $\Lambda$CDM, both models have also been constantly tested using the most up-to-date available data. Regarding the $w$CDM model, the current results that combine SN Ia, BAO, and CMB seem to show good agreement with $w=-1$~\citep{Rubin:2023ovl}. In the left panel of Fig.~\ref{fig:resultwcdm} we depict the individual contours of SN Ia, BAO, and CMB analyzes. As well as in the $\Lambda$CDM case, the results show a good overlap, indicating a concordance when $w_0\approx-1$ and $\Omega_m\approx0.3$. The results of the $w$CDM model are also presented in Tab.~\ref{tab:resultfluid}.
    
    On the other hand, using the same observational data to test the CPL parameterization has yielded results that are not in agreement with the $\Lambda$CDM limit, i.e., $w_0=-1$ and $w_a=0$. This deviation was first observed in the Union 3 analysis~\citep{Rubin:2023ovl}, as shown in Tab.~\ref{tab:resultfluid}. A similar result from the Dark Energy Spectroscopic Instrument (DESI) using BAO has been recently released. The DESI results for the CPL model have indicated a deviation from $\Lambda$CDM between $2.5\sigma$ and $3.9\sigma$, depending on the SN Ia data used. This intriguing result has triggered ongoing discussions on some critical topics: ($i$) the impact of luminous red galaxy (LRG) data on the final results from DES~\citep{Wang:2024pui}; ($ii$) the sensitivity of DESI results to the choice of statistical priors~\citep{Patel:2024odo}; and ($iii$) the potential for systematic effects in SN Ia data~\citep{Efstathiou:2024xcq}. The contour of DESI in the $w_0-w_a$ plane is shown in the right panel of Fig.~\ref{fig:resultwcdm}. The recent DESI results for the analysis of the full shape of the power spectrum confirmed the predictions made by the DESI BAO~\citep{DESI:2024jis}\footnote{Notably, this analysis has motivated important considerations for future cosmological parameter selection, particularly regarding the volume effects introduced by incorporating additional terms into power spectrum modeling.}. 

    In light of these recent findings, it is plausible to speculate that they might indicate a dynamic evolution of DE. The next years, marked by the new release from DESI data and the expectation of upcoming surveys such as Euclid, Vera Rubin, and WFIRST, will be important in confirming these results.

    \begin{table}
    \centering
    \begin{tabular}{lcccc}
    \hline\hline
    \multicolumn{5}{c}{\multirow{2}{*}{$w$CDM}}                                                                                 \\
    \multicolumn{5}{c}{}                                                                                                        \\ \hline
    Probe             & $H_0$                & $\Omega_{\rm m}$          & $w_0$                      & $w_{\rm a}$             \\ \hline
    SN Ia             & -                    & $0.244^{+0.092}_{-0.128}$ & $-0.735^{+0.169}_{-0.191}$ & -                       \\
    BAO + CMB         & $68.1^{+1.4}_{-1.3}$ & $0.308\pm0.012$           & $-0.924^{+0.044}_{-0.043}$ & -                       \\
    SN Ia + BAO + CMB & $66.6^{+0.9}_{-0.8}$ & $0.320\pm0.008$           & $-0.957^{+0.034}_{-0.035}$ & -                       \\ \hline
    \multicolumn{5}{c}{\multirow{2}{*}{CPL}}                                                                                    \\
    \multicolumn{5}{c}{}                                                                                                        \\ \hline
    BAO + CMB         & $65.0^{+2.5}_{-2.3}$ & $0.339^{+0.026}_{-0.025}$ & $-0.652^{+0.268}_{-0.261}$ & $-1.05\pm0.78$          \\
    SN Ia + BAO + CMB & $66.0\pm0.9$         & $0.329\pm0.009$           & $-0.744^{+0.100}_{-0.097}$ & $-0.79^{+0.35}_{-0.38}$ \\ \hline\hline
    \end{tabular}
    \caption{Observational results of the $w$CDM and CPL. \\ 
    \textit{Source:} Adapted from Tab. 6 of~\citet{Rubin:2023ovl}.}
    \label{tab:resultfluid}
    \end{table}

    \subsection{Interacting DE models}\label{ssec:ide}
    
    A more drastic alternative to the $\Lambda$CDM model consists of relaxing the condition that the components of the universe are independent but can interact with each other~\citep{Amendola:1999er,Billyard:2000bh,Zimdahl:2001ar}. In general, given our limited knowledge on the nature of the dark components, the interaction is assumed to involve the dark components. These models have been widely motivated as possible solutions to the cosmic coincidence problem~\citep{Chimento:2003iea}, but recently they have been used to address recent tensions~\citep{Benetti:2019lxu,DiValentino:2019jae}. 
    
    From a formal point of view, this interaction can be introduced through a non-minimal coupling with CDM~\citep{Gleyzes:2015pma}, but they can also be introduced phenomenologically via a source term in the conservation equations of the interacting components. In this case, the covariant derivatives of the energy-momentum tensors of the interacting components are no longer identically zero, but a source four-vector is introduced. Given that the universe as a whole must be conservative, the covariant derivatives of the CDM and DE must satisfy the condition
    $\nabla_{\mu}T^{\mu\nu}_{\rm c}=-\nabla_{\mu}T^{\mu\nu}_{\rm x}=Q^{\nu}$. At the background level, this condition leads to
    \begin{eqnarray}
        \dot{\rho}_{c}+3H\rho_{c}&=&Q \,, \label{eq:continuitycdm} \\
        \dot{\rho}_{\rm x}&=&-Q \,, \label{eq:continuityde} 
    \end{eqnarray}
    where $Q$ is scalar function\footnote{Note that, since CDM and DE are described as a perfect fluid, the momentum conservation is identically zero.}. 
    
    A given interacting model is explicitly defined by the particular expression chosen for the function $Q$. Despite their different dynamics, there is an explicit mapping between interacting models and models with a dynamical equation-of-state parameter, such that they result in the same Hubble rate~\citep{vonMarttens:2019ixw}. Consequently, it is not possible to distinguish between these models using distance measurement-based data. This mapping may extend to the perturbative level, depending upon the sound speed of DE.
    
    \section{Conclusions}\label{sec:conclusions}
    
    In this chapter, we have presented some of the main aspects of the discussion on the accelerated expansion of the Universe. First, we made a brief overview of the main developments culminating in the direct observation of cosmic acceleration through the analysis of SN Ia. Furthermore, we established the mathematical framework of the standard cosmological model, where the accelerated expansion is driven by the cosmological constant. In the context of the $\Lambda$CDM model, we have also discussed some of its main aspects, including observational successes, open questions, and recent cosmological tensions. 
    
    Beyond the context of the $\Lambda$CDM model, we have also addressed some of the main alternatives to describe DE as an extra exotic fluid. More precisely, we focused on the quintessence and $k$-essence models as a dynamical DE component and also models where DE is allowed to interact with CDM. 
    
    Looking ahead, the forthcoming observational data will be extremely important in refining our understanding of these complex issues.

    \section*{Acknowledgments}

    It is a pleasure to thank Winfried Zindahl, Valerio Marra, and Uendert Andrade for their critical reading and insightful comments.
    RvM is suported by Fundação de Amparo à Pesquisa do Estado da Bahia (FAPESB) grant TO APP0039/2023.
    JA is supported by CNPq grant No. 307683/2022-2 and Fundação de Amparo à Pesquisa do Estado do Rio de Janeiro (FAPERJ) grant No. 259610 (2021).

    \bibliographystyle{Harvard}
    \bibliography{reference}

\end{document}